\documentclass[fleqn,usenatbib]{mnras}
\pdfoutput=1

\usepackage[T1]{fontenc}
\usepackage{ae,aecompl}
\usepackage{color, soul}
\usepackage{verbatim}
\usepackage{units}
\usepackage{subfig}
\usepackage{pdflscape}
\usepackage{svg}
\usepackage{bm}
\usepackage{stfloats}
\usepackage{float}
\usepackage{multicol}
\usepackage{hyperref}
\usepackage{multirow}

\usepackage{graphicx}	
\usepackage{amsmath}	
\usepackage{amssymb}	
\usepackage{epstopdf}
\usepackage[normalem]{ulem} 

\defcitealias{Lu2013}{L13}
\defcitealias{Bartko2010}{B10}
\defcitealias{Salpeter1955}{S55}

\newcommand\clearrow{\global\let\rowmac\relax}
\newcommand{\cmmnt}[1]{\ignorespaces}

\title[\textit{Gaia} HVSs and the GC]{Constraints on the Galactic Centre environment from \textit{Gaia} hypervelocity stars}

\author[Evans et al.]{
F. A. Evans$^{1}$\thanks{E-mail: evans@strw.leidenuniv.nl},
T. Marchetti$^{2}$,
E. M. Rossi$^{1}$ \\
$^{1}$Leiden Observatory, Leiden University, PO Box 9513, NL-2300 RA Leiden, The Netherlands\\
$^{2}$European Southern Observatory, Karl-Schwarzschild-Strasse 2, 85748 Garching bei M{\"u}nchen, Germany \\
}

\date{Accepted XXX. Received YYY; in original form ZZZ}

\pubyear{2021}

\begin{document}
\label{firstpage}
\pagerange{\pageref{firstpage}--\pageref{lastpage}}
\maketitle

\begin{abstract}
Following a dynamical encounter with Sgr A*, binaries in the Galactic Centre (GC) can be tidally separated and one member star ejected as a hyper-velocity star (HVS) with a velocity beyond the escape speed of the Milky Way. As GC-born objects located in more observationally accessible regions of the sky, HVSs offer insight into the stellar population in the inner parsecs of the Milky Way. We perform a suite of simulations ejecting stars from the GC, exploring how detectable HVS populations depend on assumptions concerning the GC stellar population, focusing on \textcolor{black}{HVSs} which would appear in current and/or future data releases from the \textit{Gaia} space mission with precise astrometry and measured radial velocities. We show that predictions are sensitive to two parameters in particular: the shape of the stellar initial mass function (IMF) in the GC and the ejection rate of HVSs. The absence of confident HVS candidates in \textit{Gaia} Data Release 2 excludes scenarios in which the HVS ejection rate is $\gtrsim3\times10^{-2} \, \mathrm{yr^{-1}}$. Stricter constraints will be placed on these parameters when more HVS candidates are unearthed in future \textit{Gaia} data releases -- assuming recent determinations of the GC IMF shape, one confident HVS \textit{at minimum} is expected in \textit{Gaia} DR3 and DR4 as long as the HVS ejection rate is greater than $\sim 10^{-3} \, \mathrm{yr^{-1}}$ and $\sim10^{-5} \, \mathrm{yr^{-1}}$, respectively. 
\end{abstract}




\section{Introduction}

The central parsec of the Milky Way is known to contain a population of Wolf-Rayet/O- and B-type stars \citep{Allen1990, Krabbe1991, Krabbe1995, Ghez2003, Lu2009, Bartko2010, Do2013, Feldmeier2015}. The existence of this young stellar cluster (YSC) in this region is perplexing, as molecular clouds should not survive the strong tidal forces from Sgr A*, the $\sim$4$\times10^{6} \, \mathrm{M_\odot}$ supermassive black hole (SMBH) at the centre of the Galaxy \citep{Eisenhauer2005,Ghez2008,Genzel2010}. Of particular interest is the central arcsecond ($\sim$0.04$ \, \mathrm{pc}$) of the Galactic Centre (GC), where the so-called S-star cluster can be found, a population of B-dwarfs on isotropically distributed, eccentric orbits about Sgr A* \citep[e.g.][]{Gillessen2009, Gillessen2017}. An appealing explanation for the origin of the S-star cluster is the scattering \textcolor{black}{or diffusion} of binaries from the YSC -- or the more extended nuclear star cluster \citep[NSC; see][]{Boker2010} in which it resides -- \textcolor{black}{onto orbits which bring them close to} Sgr A* \citep{Lightman1977, Perets2007, Madigan2009, Madigan2014, Generozov2020, Generozov2021}. Via the so-called Hills mechanism \citep{Hills1988}, the binary is tidally disrupted following a close encounter with Sgr A*: one member of the former binary is ejected at a high velocity while the other remains bound to Sgr A* as an S-star \citep[e.g.][]{Gould2003,Yu2003,Bromley2006, Ginsburg2006, Zhang2013}.

The characterisation of the properties and dynamics of the YSC (and NSC in general) is crucial to obtaining an understanding of star formation under extreme conditions, the interplay between Sgr A* and its environs \citep[see][for a review]{Genzel2010} and the growth of Sgr A* \citep[e.g.][]{Bromley2012}. The existence of nuclear star clusters in other galaxies indicates this feature is not isolated to the Milky Way \citep[see][for a review]{Neumayer2020}. The Galactic NSC is therefore a closeby and valuable laboratory for investigating this general ingredient in the structure of galaxies.

Direct observation of the GC region, however, is beset by challenges. Source-crowding along the line of sight to the GC is extreme and the extinction and reddening of light by interstellar dust is severe and inhomogeneous \citep[see][for a review]{Schodel2014}. The ejected former companions of the S-stars can be a convenient complementary tool to circumvent these hurdles. Following the tidal disruption of the binary, these so-called hyper-velocity stars (HVSs) are ejected at a characteristic velocity of $\sim$1000$ \, \mathrm{km \ s^{-1}}$ \citep{Hills1988}, in excess of the Galactic escape speed. With such high velocities, HVSs easily escape the GC region and can be found all across the sky. As former members of the NSC spotted in more observationally accessible regions of the sky, HVSs can then be used as a clever `back-door' tracer to investigate stellar populations in the GC. Since the first serendipitous HVS candidate detections \citep{Brown2005, Hirsch2005, Edelmann2005}, dozens of candidates have been uncovered via targeted searches \citep{Brown2006, Brown2009, Brown2012, Brown2014}, revisits of curious objects \citep[e.g][]{Heber2008,Tillich2009,Irrgang2010,Irrgang2019} and searches within the archives of ongoing large Galactic surveys \citep[e.g.][]{Zhong2014, Huang2017, Li2018, Hattori2018, Marchetti2019, Koposov2020, Li2021, Marchetti2021}. See \citet{Brown2015rev} for a review of these objects and the Open Fast Stars Catalog\footnote[1]{https://faststars.space/} \citep{Guillochon2017,Boubert2018} for an up-to-date list of known fast stars.

While the potential of HVSs to trace the Galactic potential has generated interest \citep{Gnedin2005,Yu2007,Kenyon2008,Kenyon2014,Rossi2017,Contigiani2019,Hattori2019}, their ability to constrain the GC stellar population has been comparatively less-explored. Since the ejection velocity of a star ejected via the Hills mechanism depends explicitly on the properties of the disrupted progenitor binary \citep{Sari2010, Kobayashi2012, Rossi2014}, the size, properties and kinematics of detected HVS populations can provide constraints on the stellar initial mass function (IMF) in the GC, the distributions of orbital periods and mass ratios among GC binaries, and the still-uncertain ejection rate of HVSs from the GC \citep[see][]{Brown2015rev}. Using the total velocities of 21 HVSs with Galactocentric total velocities $>275 \, \mathrm{km \ s^{-1}}$ from the MMT HVS Survey \citep[see][]{Brown2014}, \citet{Rossi2017} provided joint constraints on the Galactic potential and the GC stellar environment (c.f. their Fig. 2). Their results point to an escape speed from the GC to $50 \, \mathrm{kpc}$ of $\lesssim850 \, \mathrm{km \ s^{-1}}$ and generally prefer a steeply declining distribution of binary mass ratios.

In this work we complement the work of \citet{Rossi2017} by considering current and future data releases from the European Space Agency's \textit{Gaia} mission \citep{Gaia2016,Gaia2018,Gaia2021}, which aims to acquire five-parameter astrometric solutions (position, parallax, proper motion) for billions of Milky Way objects and radial velocity measurements for a subset of tens of millions of bright sources. While \textit{Gaia} has been instrumental in refining the kinematics of known HVS candidates \citep{Irrgang2018, Brown2018, Erkal2019,Kreuzer2020} and in recategorizing many late-type HVS candidates as actually bound to the Galaxy \citep{Boubert2018}, it has not yet proven to be a bountiful wellspring of new promising HVS candidates. Several stars with high-quality astrometry and high probabilities ($>80$\%) of being unbound to the Galaxy have been discovered both in \textit{Gaia} Data Release 2 \citep[DR2;][]{Bromley2018, Hattori2018, Marchetti2019} and Early Data Release 3 \citep[EDR3;][]{Marchetti2021}, however, none can be confidently associated with an origin in the Galactic Centre. This is due in large part to the fact that identifying HVSs as being unbound requires a determination of 3D velocities, and \textit{Gaia} DR2 and EDR3 provide measured radial velocities only for the subset of $\sim7$ million sources which are brighter than the 12th magnitude in the \textit{Gaia} Radial Velocity Spectrometer (RVS) passband.

While disappointing at first glance, \textcolor{black}{we show in this work that} this paucity of high-confidence HVSs in the \textit{Gaia} DR2 and EDR3 radial velocity catalogues is both a) expected, and b) a valuable piece of observational evidence in and of itself. Given sufficient knowledge of the \textit{Gaia} DR2/EDR3 selection function, i.e. an understanding of which stars \textcolor{black}{do and do not} appear in \textit{Gaia} DR2/EDR3 \citep[see][]{Boubert2020cogi,Boubert2020cogii,Rybizki2021,Everall2021cogv}, an absence of HVSs is sufficient to exclude robustly regions of the ejection rate vs. GC IMF shape parameter space. Moreover, prospects are more promising for the full \textit{Gaia} Data Release 3 (DR3, expected Q2 2022) and the fourth and final data release of the nominal mission (DR4), which will provide more precise astrometry and radial velocity measurements for fainter objects. We can make testable predictions of the HVS populations which will lurk in these upcoming data releases. When future HVS populations are unearthed, they can be used to place more strict constraints on the stellar populations in the GC. In this work we explore the sensitivity of the \textit{Gaia} HVS populations to variations of the NSC  initial mass function (IMF), the orbital period and mass ratio distribution among NSC binaries, and the ejection rate of HVSs from the GC to help rule out models of the GC stellar environment which predict an over- or under-abundance of HVSs. 

This paper is organized as follows. In Sec. \ref{sec:methods} we describe our HVS ejection model in detail. In Sec. \ref{sec:constraints} we briefly review contemporary constraints from the literature on the GC stellar population. In Sec. \ref{sec:results} we present our results. Finally we discuss these results and present our conclusions in Sec. \ref{sec:discussion}.

\section{Hills Ejection Model} \label{sec:methods}

In the following subsections we describe the various ingredients of our HVS ejection model. We describe how we draw the \textcolor{black}{ejection velocities and masses of stars ejected via the Hills mechanism}, how we assign ages and flight times to ejected stars, how we propagate ejected stars forward in time, how we obtain mock \textit{Gaia} photometric and astrometric observations for these populations, and how we identify mock stars which would stand out as promising HVS candidates in current and future \textit{Gaia} data releases.

\subsection{\textcolor{black}{Ejection} masses and velocities}
\label{sec:methods:ejection}

\begin{table*}
\begin{tabular}{lllll}
Parameter & Variable & Distribution & Fiducial value & Range  \\ \hline

Initial mass function $f(m)$ & $\kappa$ & $f(m) \propto m^{\kappa}$ & $\kappa=-1.7$ & $\kappa = [-3.3,-0.3]$ \\
Binary orbital period $f(P)$ (days) & $\pi$ & $f(\mathrm{log}P)\propto (\mathrm{log}P)^{\pi}$ & $\pi=0$ & $\pi=[-2,+2]$ \\
Binary mass ratio $f(q\equiv m_2 / m_1)$ & $\gamma$ & $f(q) \propto q^{\gamma}$ & $\gamma=-1$ & $\gamma=[-3,+2]$\\
HVS ejection rate $\eta$ ($\mathrm{yr^{-1}}$) & $\eta$ & $n/a$ & $\eta=10^{-4}$ & $\eta=[10^{-6},10^{-1}]$ \\

\end{tabular}
\caption{Summary of the GC parameters we vary in our ejection models and attempt to constrain in this work: the physical parameter (Col. 1), the corresponding variable (Col. 2), the probability distribution function of the parameter (Col. 3), and the assumed fiducial value and full domain we explore for the variable (Cols. 4, 5). See text for details.}
\label{tab:vars}
\end{table*}

We generate populations of Hills mechanism-ejected stars from the GC following a Monte Carlo (MC) ejection model similar to that implemented in \citet{Evans2021} \citep[see also][]{Marchetti2018}. We describe the model as follows.

We generate binaries defined by three parameters: $m_{\rm p}$, the zero-age main sequence (ZAMS) mass of the primary; $q\equiv m_{\rm s}/m_{\rm p}$, the ZAMS mass ratio between the secondary ($m_{\rm s}$) and primary; and $a$, the orbital semi-major axis \textcolor{black}{of the stars within the binary. An $m_{\rm p}$ is drawn for each binary in the range [0.1, 100] $\mathrm{M_\odot}$ assuming a power-law initial mass function, i.e. $f(m_{\rm p})\propto m_{\rm p}^{\kappa}$. We choose $\kappa=-1.7$ as our fiducial GC IMF slope following \citet{Lu2013}. Mass ratios are sampled in the interval $0.1\leq q \leq 1 $ distributed also as a power law, $f(q)\propto q^{\gamma}$, with a fiducial value for $\gamma$ of -1 \citep[e.g.][]{Moe2013, Sana2013, Moe2015}. Finally, we assign to each binary an orbital semi-major axis assuming its orbital period about its centre of mass is distributed as $f(\mathrm{log}P)\propto (\mathrm{log}P)^{\pi}$. We assume $\pi=0$ as our fiducial value \citep[e.g.][]{Moe2013, Kobulnicky2014, Dunstall2015, Almeida2017}. Maximum and minimum periods are set such that the binary semi-major axis $a$ is constrained to the range $2.5\, \text{max}(R_{\rm p}, R_{\rm s})\leq a \leq 2000 \, \mathrm{R_\odot}$, where $R_{\rm p}$ and $R_{\rm s}$ are the stellar radii of the primary and secondary stars, respectively. This lower limit is set by Roche lobe overflow. In Table \ref{tab:vars} we summarize these distributions and fiducial assumptions. We also include the ranges of $\kappa$, $\gamma$ and $\pi$ we explore in our model variations. In Sec. 3 we discuss these assumptions in greater depth and justify our chosen ranges for $\kappa$, $\gamma$ and $\pi$.} 

\textcolor{black}{We assume that two-body relaxation processes cause a binary in the GC environment with a total mass $M$ and orbital separation $a$ to diffuse in phase space until its angular momentum reaches $J\simeq2GM r_{\rm t}$, sufficient for the binary to approach Sgr A* within its tidal radius $r_{\rm t}\equiv a(3M_{\rm Sgr A^*}/M)^{1/3}$ and be tidally separated. In this regime, the separation probability only weakly on the binary tidal radius but instead it is largely set by the relaxation timescale \citep{Lightman1977, Alexander2005, Rossi2014}.} Following the binary disruption, one member star is ejected while the other remains bound to the MBH. We assume here that the progenitor binary approaches Sgr A* on a parabolic orbit -- in this situation, there is an equal probability for ejecting either star in the binary \citep{Sari2010,Kobayashi2012}. We therefore randomly designate one star among the binary as the ejected star. Its ejection velocity is calculated analytically \citep{Sari2010, Kobayashi2012,Rossi2014}:
\begin{equation}
    v_{\rm ej} = \sqrt{\frac{2Gm_{\rm c}}{a}} \left( \frac{M_{\rm Sgr A^*}}{M} \right)^{1/6} \, \text{,}
    \label{eq:vej}
\end{equation}
where $M=(1+q)m_{\rm p}$ is the total system mass of the progenitor binary, $m_{\rm c}$ is the mass of the companion that remains bound to Sgr A*, and $M_{\rm Sgr A^*}=4\times10^{6} \, \mathrm{M_\odot}$ is the mass of Sgr A* \citep{Eisenhauer2005, Ghez2008}. \textcolor{black}{In our analytical approach, we omit from Eq. \ref{eq:vej} a multiplicative factor which depends on the HVS progenitor binary phase, orbital inclination, tidal radius and orbital pericentre to Sgr A*, as it averages out to unity when considering all orbital geometries \citep{Sari2010}. However, a full numerical calculation of this factor may influence the velocity distribution of the fastest stars \cite[see][c.f. Sec. 6]{Rossi2014}}.

\begin{figure*}
    \centering
    \includegraphics[width=0.65\columnwidth]{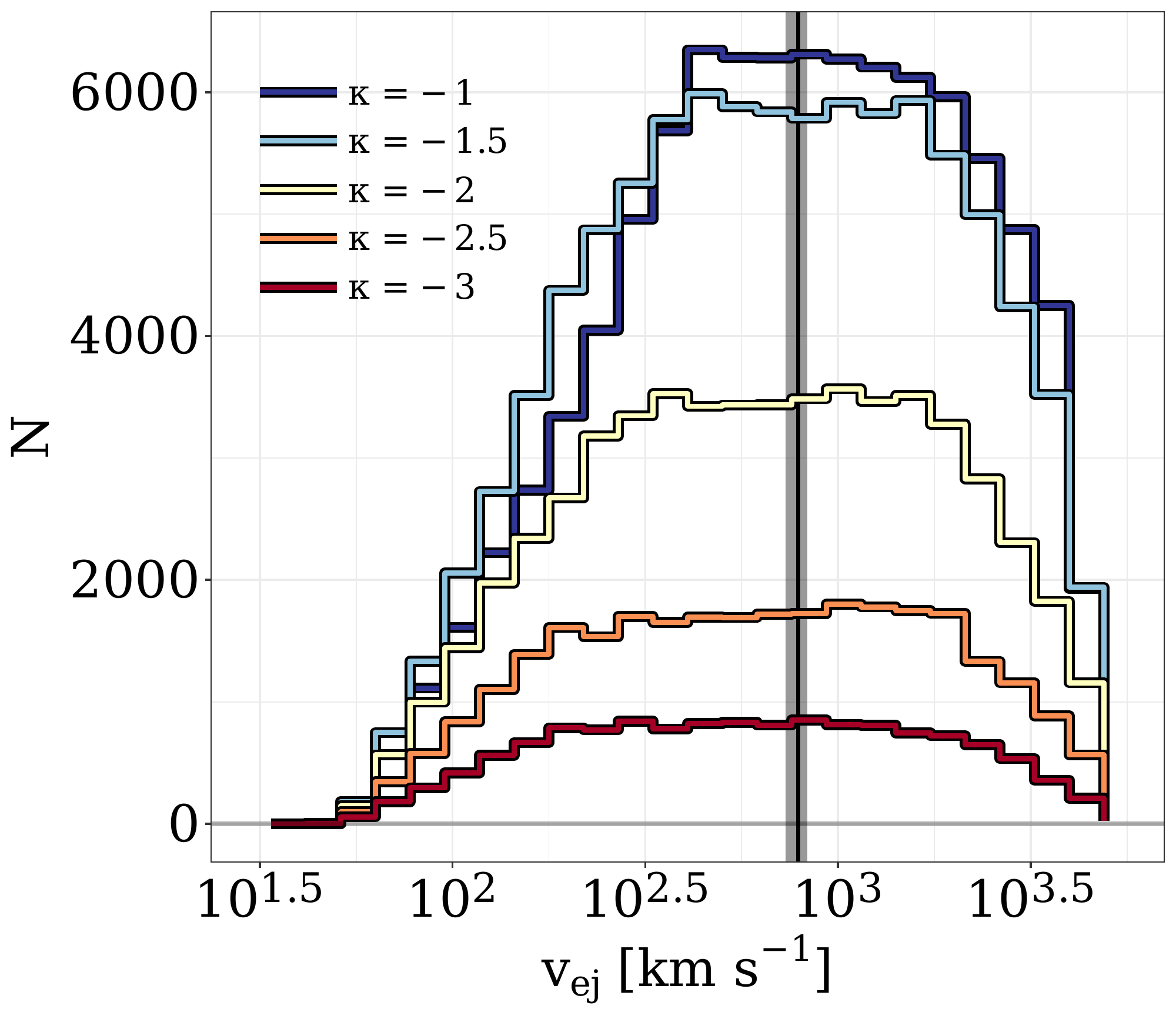}
    \includegraphics[width=0.65\columnwidth]{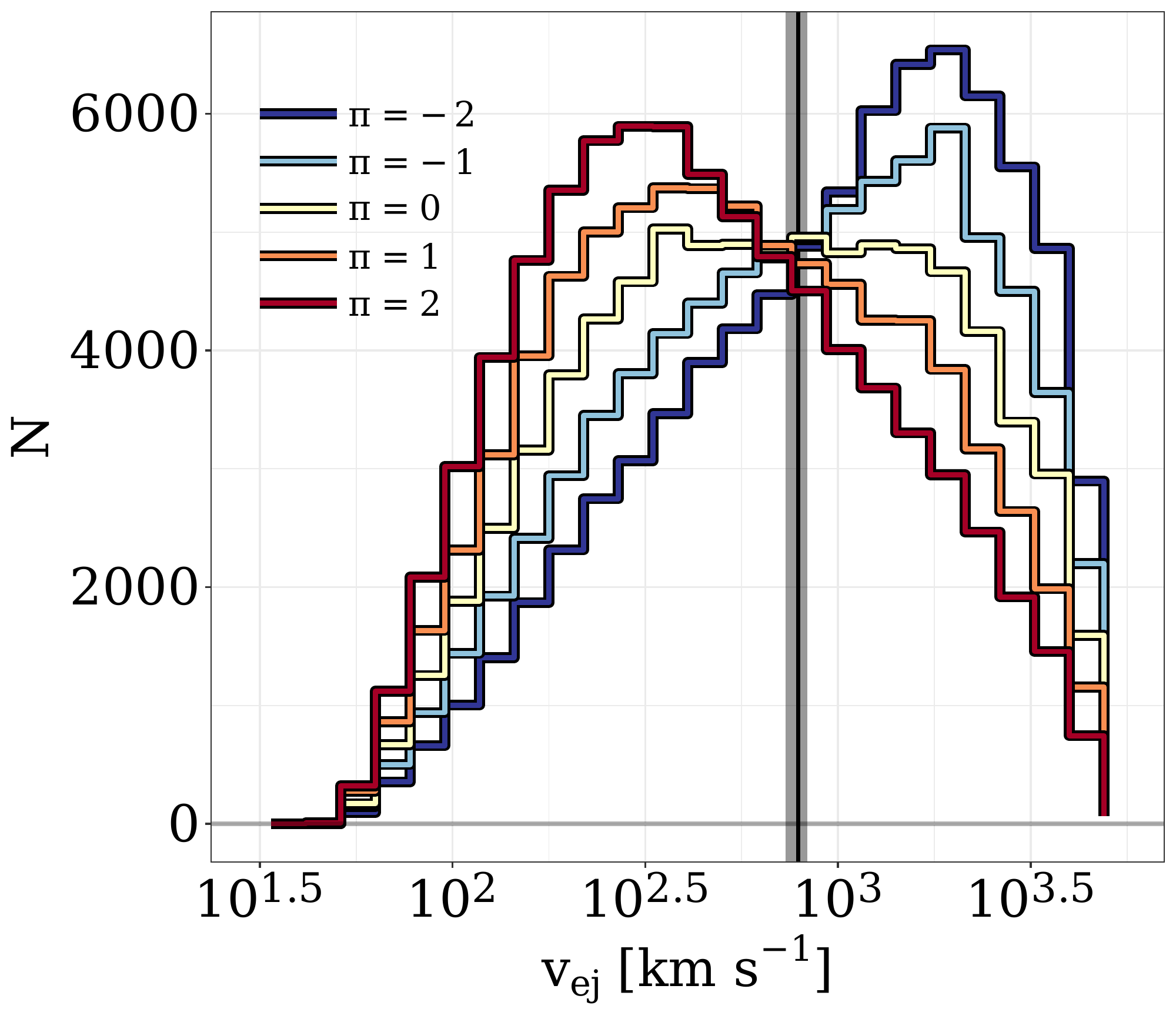}
    \includegraphics[width=0.65\columnwidth]{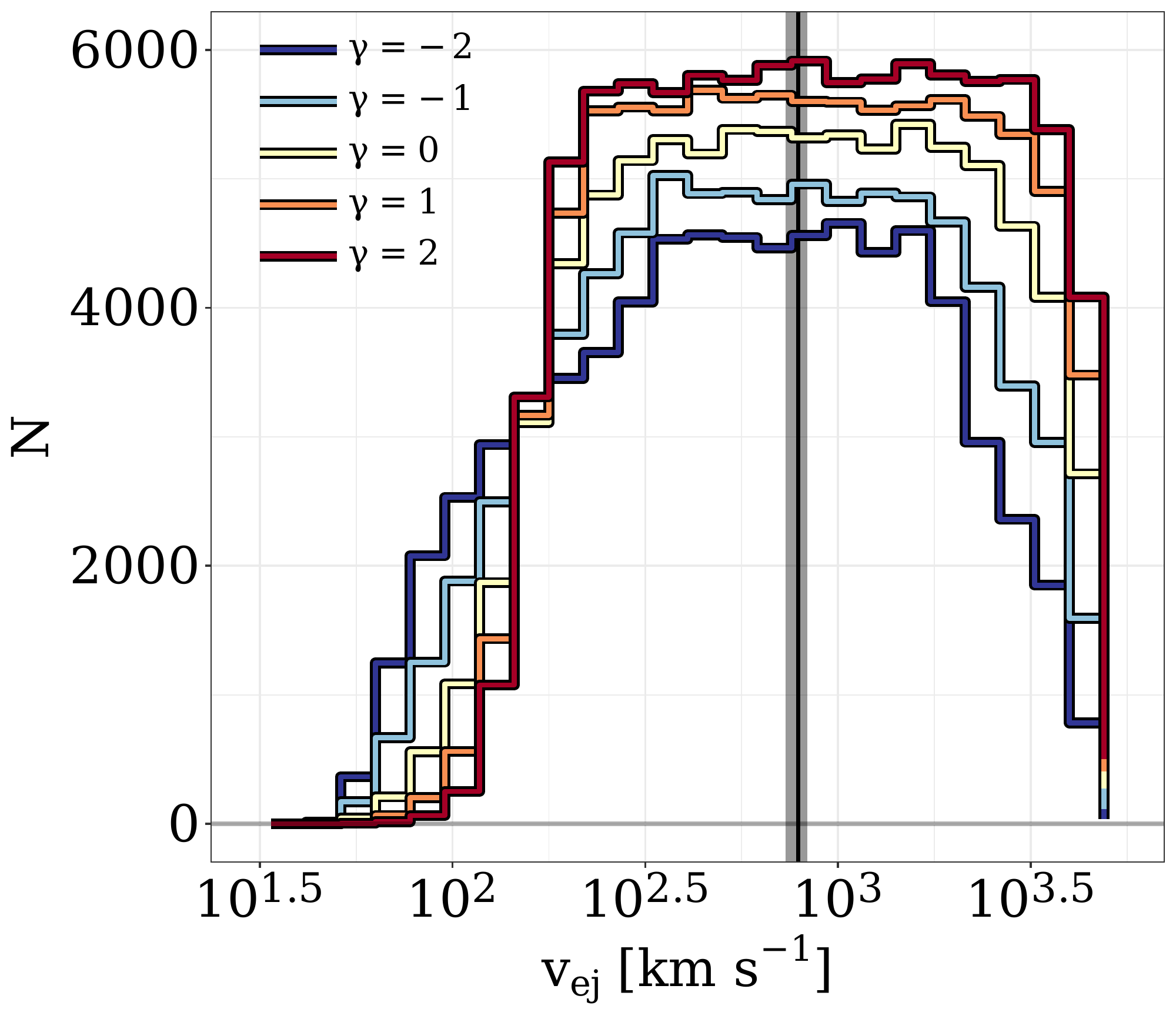}
    \caption{The initial velocities of $m>0.5 \, \mathrm{M_\odot}$ stars ejected via the Hills mechanism (see Sec. \ref{sec:methods}). Panels show how the $v_{\rm ej}$ distribution changes with the initial mass function power law index $\kappa$ of stars in the NSC (left), the orbital log-period power law index $\pi$ of progenitor binaries (middle) and the binary mass ratio power law index $\gamma$ (right). The vertical line and shaded band show $v_{\rm ej}=789_{-39}^{+43} \, \mathrm{km \ s^{-1}}$, the typical escape velocity from $3 \, \mathrm{pc}$ to infinity in the family of Galactic potentials described in Sec. \ref{sec:methods:potential}. In panels where they are not being varied, $\kappa$, $\pi$ and $\gamma$ are set to their fiducial values (see Table \ref{tab:vars})}.
    \label{fig:vejs}
\end{figure*}

\textcolor{black}{We illustrate in Fig. \ref{fig:vejs} how the distributions of ejection velocities depend individually on $\kappa$, $\gamma$ and $\pi$.} We show $v_{\rm ej}$ histograms only for $m>0.5 \, \mathrm{M_\odot}$ ejected stars, as stars less massive than $0.5 \, \mathrm{M_\odot}$ ejected from the GC are extremely unlikely to end up bright enough to appear in current or future \textit{Gaia} data releases with measured radial velocities. In the left panel of Fig. \ref{fig:vejs} we see unsurprisingly that the number of $m>0.5 \, \mathrm{M_\odot}$ ejected stars is highly dependent on the IMF index \textcolor{black}{-- as $\kappa$ becomes more negative, a smaller proportion of stars have initial masses greater than $0.5 \, \mathrm{M_\odot}$} Among the $m>0.5 \, \mathrm{M_\odot}$ stars, however, the $v_{\rm ej}$ distribution remains consistent. In the middle panel $v_{\rm ej}$ depends strongly on the binary log-period distribution index $\pi$ -- a lower $\pi$ means HVS progenitor binaries are tighter on average, therefore orbital velocities are higher at the moment of binary \textcolor{black}{tidal separation}. The vertical line shows $v_{\rm ej}=789_{-39}^{+43} \, \mathrm{km \ s^{-1}}$, the average escape velocity from the GC to infinity in the family of Milky Way gravitational potentials we employ (see Sec. \ref{sec:methods:potential}). 58\% of $m>0.5 \, \mathrm{M_\odot}$ stars are ejected above this velocity for $\pi=+2$ while only 28\% of $m>0.5 \, \mathrm{M_\odot}$ stars satisfy this for $\pi=-2$.  Finally, in the right panel we see that the number of $m>0.5 \, \mathrm{M_\odot}$ stars increases with an increasing $\gamma$, \textcolor{black}{since a greater proportion of ejected stars will satisfy $m>0.5 \, \mathrm{M_\odot}$ when $\gamma$ is large.} The $v_{\rm ej}$ distribution remains otherwise fairly unchanged.

\subsection{HVS ejection rate and flight time distribution} \label{sec:methods:tflight}

In this work we consider only ejections over the past $100 \, \mathrm{Myr}$, as any HVS unbound to the Galaxy ejected more than $100 \, \mathrm{Myr}$ ago will currently be too far (and thus too faint) to appear in the \textit{Gaia} DR4 radial velocity catalogue. Stars are ejected from the GC with a rate $\eta$ (see Sec. \ref{sec:constraints}), assumed here to be constant \textcolor{black}{with a fiducial value of $\eta=10^{-4} \, \mathrm{yr^{-1}}$ (see Table \ref{tab:vars})}. For each ejected star we assign a flight time and a stellar age at ejection.  Our present day mock catalogue consists of stars ejected $t_{\rm ej}$ ago that have not yet left the main sequence. We assign $t_{\rm ej}$ uniformly: 
\begin{align}
    t_{\rm ej} &= \epsilon_1 \cdot \mathrm{100 \, Myr}  \; ,
\label{eq:tflight}
\end{align}
where $0<\epsilon_1 < 1$ is a uniformly distributed random number. We assume there is no preferred stellar age at the time of ejection. The age of a star at ejection $t_{\rm age, ej}$ is then a random fraction $\epsilon_2$ of its main sequence lifetime $t_{\rm MS}$;
\begin{equation}
    t_{\rm age, ej} = \epsilon_2 \cdot t_{\rm MS} \; ,
\end{equation}
where $t_{\rm MS}$ is calculated according to the stellar mass and metallicity using the \citet{Hurley2000} analytical formulae. The mean stellar metallicity in the GC is slightly super-solar with a large spread \citep{Do2015, FeldmeierKrause2017, Rich2017, FeldmeierKrause2020, Schodel2020}. We model this large spread by assigning each ejected star a metallicity $\xi \equiv \mathrm{log_{10}}[Z/Z_\odot]$ of -0.5, 0.0 or +0.5 with equal probability. The solar metallicity is assumed here to be $Z_{\odot}=0.02$ \citep{Anders1989}.

The remaining main sequence lifetime of the star $t_{\rm left}$ is
\begin{equation}
    t_{\rm left} = t_{\rm MS} - t_{\rm age, ej} = (1-\epsilon_2) \cdot t_{\rm MS} \; .
\end{equation}
We remove stars for whom $t_{\rm ej}>t_{\rm left}$. The flight time of each mock ejected star is then
\begin{equation}
    t_{\rm flight} = t_{\rm ej}
\end{equation}
and its current age is
\begin{equation} \label{eq:tage}
    t_{\rm age,0} = t_{\rm age,ej} + t_{\rm flight} \; .
\end{equation}
These stellar ages are used later in obtaining mock \textit{Gaia} photometry (see Sec. \ref{sec:methods:observations}).

After assigning masses, velocities, flight times and ages according to Eqs. \ref{eq:vej}-\ref{eq:tage} and removing ejected stars which do not survive on the main sequence until the present day, we are left with $\sim$78000 ejected stars for our fiducial model.

\subsection{Orbital integration} \label{sec:methods:potential}

We propagate Hills mechanism-ejected stars through a family of realistic Milky Way potentials described by \citet{McMillan2017}. We describe briefly the potentials here and refer to \citet{McMillan2017} for more information. The potentials consist of an axisymmeterized approximation of a \citet{Bissantz2002} bulge, H$\textsc{I}$ and molecular hydrogen discs each modeled as exponential discs with central holes following \citet{Dehnen1998}, exponential discs for both the thin and thick stellar discs, and a spherical NFW dark matter halo \citep{Navarro1996}. While the gas discs remain fixed, the parameters of the remaining components are left as free parameters and are fit to various kinematic data using a Monte Carlo Markov Chain (MCMC) method. This method additionally provides estimates for the distance $R_0$ from the Sun to the GC, the circular velocity $v_{\rm 0}$ of the Galaxy at $R_{\rm0}$, and the 3D peculiar velocity $v_\odot\equiv(U_\odot,V_\odot,W_\odot)$ of the Sun with respect to a circular orbit at $R_0$.

We use the entire \citet{McMillan2017} MCMC chain (P. McMillan, private communication). For each realization in which we eject and propagate a population of stars \textcolor{black}{ejected via the Hills mechanism}, we draw a MW potential and solar position/velocity from among the post-burn-in entries in the chain. By averaging our results over many iterations, we marginalize our predictions over contemporary uncertainties in the Galactic potential.

We implement drawn potentials and integrate ejected star orbits using the \texttt{PYTHON} package \texttt{GALPY}\footnote[2]{\url{https://github.com/jobovy/galpy}} \citep{Bovy2015}. We assume the distribution of ejection directions is isotropic with respect to the GC. Ejected stars are therefore initialized at random points on the surface of a sphere $3 \, \mathrm{pc}$ in radius centred on Sgr A$^{*}$. This is the so-called `sphere of influence', where the contribution of Sgr A$^{*}$ to the total Milky Way potential becomes subdominant \citep{Genzel2010}. Stars are ejected with initial velocities pointing directly away from the GC. Each star is integrated forward in time for its flight time through the drawn Galactic potential. Orbits are integrated using a fifth-order Dormand-Prince integrator \citep{Dormand1980} with a fixed timestep of $0.1 \, \mathrm{Myr}$. 

\subsection{Mock \textit{Gaia} photometry} \label{sec:methods:observations}

After propagating each ejected star for its appropriate flight time, we estimate its photometric properties following \citet{Marchetti2018}. We first compute each star's radius, effective temperature and surface gravity using the analytic formulae of \citet{Hurley2000} given its mass, metallicity and present age. We find the best-fitting spectrum among the $\xi=-0.5$, $\xi=0.0$ or $\xi=+0.5$ spectra in the BaSeL SED Library 3.1 \citep{Westera2002} via chi-squared minimization of the effective temperature and surface gravity. We assume an atmospheric micro-turbulence velocity of $2 \ \mathrm{km \, s^{-1}}$ and a mixing length of zero. We calculate the visual extinction $A_{\rm V}$ at each star's position using the \texttt{MWDUST}\footnote[3]{\url{https://github.com/jobovy/mwdust}} three-dimensional Galactic dust map \citep{Bovy2016}, which is itself a combination of the \citet{Drimmel2003}, \citet{Marshall2006}, and \citet{Green2015} maps. We calculate the attenuation $A_{\lambda}$ at all wavelengths assuming a \citet{Cardelli1989} reddening law with $R_{\rm V}=3.1$. We estimate the star's apparent magnitude in the \textit{Gaia G} band by integrating the flux of the best-fitting BaSeL spectrum and the attenuation $A_{\lambda}$ through the \textit{Gaia} Early Data Release 3 $G$ passband\footnote[4]{\url{https://www.cosmos.esa.int/web/gaia/edr3-passbands}} \citep[see][eq. 1]{Jordi2010}. We compute as well apparent magnitudes in the Johnson-Cousins \textit{I$_{C}$} and \textit{V} bands, adopting the \citet{Bessell1990} passbands. The apparent magnitude in the \textit{Gaia G$_{\rm RVS}$} band can then be computed using from these the polynomial fits in \citet{Jordi2010} (table 3). 

\subsection{The \textit{Gaia} \textcolor{black}{spectroscopic} selection function}

We describe here how we determine which ejected stars would appear in \textit{Gaia} DR2, DR3 or DR4 with an assigned radial velocity. Only a small fraction of our ejected stars would be bright enough to appear in the \textit{Gaia} source catalogue, and an even smaller proportion would satisfy the brightness and effective temperature criteria to be assigned a validated radial velocity.  

Radial velocities are published in \textit{Gaia} DR2 \citep{Gaia2018} for $\sim$7 million sources brighter than $G_{\rm RVS}\approx12$ mag in the effective temperature range $[3500\, \mathrm{K}, 6900 \, \mathrm{K}]$ \citep{Katz2019}. This upper limit roughly corresponds to stars of spectral type early-F to late-A -- significant blending of Paschen lines with the calcium triplet prevents the \textit{Gaia} DR2 spectroscopic pipeline from obtaining validated radial velocities for hotter stars. Note, however, that the \textit{Gaia} DR2 radial velocity catalogue is not 100\% complete down to $G_{\rm RVS}=12$, nor is it entirely incomplete beyond. Since the \textit{Gaia} astrometric and spectroscopic analysis pipelines require a source to be detected a minimum number of times to be assigned a full astrometric solution and radial velocity, the precise \textit{Gaia} spectroscopic selection function will be impacted by the spinning-and-precessing scanning law of the \textit{Gaia} satellite and by stellar crowding in dense regions such as the GC \citep[see][]{Boubert2020cogi, Boubert2020cogii}. \textcolor{black}{We use the \textit{Gaia} DR2 spectroscopic selection function as estimated by \citet{Everall2021cogv}\footnote[5]{see \url{https://github.com/gaiaverse/selectionfunctions}} to identify which mock ejected stars would be included in \textit{Gaia} DR2 radial velocity catalogue.} We designate a star as included if a randomly drawn random number $0<\epsilon<1$ satisfies $\epsilon<p$, where $p$ is a detection probability depending on the position, apparent $G$-band magnitude and $G-G_{\rm RP}$ colour of the mock ejected star.

The third \textit{Gaia} data release has been split into two parts. \textit{Gaia} Early Data Release 3 parallax precisions improve by 30 per cent relative to DR2, and proper motion precisions by a factor of 2 \citep{Gaia2020EDR3}. New or updated radial velocity measurements are not included in \textit{Gaia} EDR3 -- radial velocity measurements of DR2 sources have been ported to their EDR3 counterparts. The full \textit{Gaia} DR3 (expected Q2 2022) \textcolor{black}{includes the EDR3 astrometric solutions as well as new and updated radial velocity measurement}s. Radial velocities will be available for $\sim$33 million sources brighter than $G_{\rm RVS}\simeq14$ in the temperature range effective temperature range $[3500\, \mathrm{K}, 6900 \, \mathrm{K}]$ \citep{Katz2019}. Improvements in the \textit{Gaia} spectroscopic pipeline\footnote[6]{see \url{https://www.cosmos.esa.int/web/gaia/dr3}} will additionally allow validated radial velocity measurements for $7000 \, \mathrm{K} < T_{\rm eff} < 14500 \, \mathrm{K}$ sources to a depth of $G_{\rm RVS}\lesssim12$. In the absence of more detailed information about the DR3 spectroscopic selection function, we select HVSs \textcolor{black}{detectable in the DR3 radial velocity catalogue} using solely these magnitude and effective temperature criteria. \textcolor{black}{We do the same for DR4 --} radial velocities in DR4 will be available for sources cooler than $6900 \, \mathrm{K}$ down to the $G_{\rm RVS}=16.2$ mag limiting magnitude of the RVS spectrometer \citep{Cropper2018, Katz2019}, and for hotter stars radial velocities will be available for sources brighter than $G_{\rm RVS}=14$. 

\subsection{\textcolor{black}{Mock \textit{Gaia} astrometric \& radial velocity uncertainties}} \label{sec:methods:errors}

\textcolor{black}{In order to select stars which would be easily identifiable as promising HVS candidates in current and future \textit{Gaia} data releases, it is useful to  estimate each star's DR2, (E)DR3 and DR4 astrometric and radial velocity uncertainties. For DR2 astrometric uncertainties we use the DR2 astrometry spread function of \citet{Everall2021cogiv}\footnote[7]{see \url{https://github.com/gaiaverse/scanninglaw}}, which computes the full 5D astrometric covariance matrix for a source based on its sky position and \textit{Gaia G}-band magnitude. The covariance matrix provides the uncertainties on a star's astrometric solution (parallax, position, proper motion) as well as the correlations among these. We similarly estimate (E)DR3 astrometric uncertainties using the \citet{Everall2021cogiv} EDR3 astrometry spread function. We estimate the DR4 astrometric covariance matrix for each star by taking the EDR3 covariance matrix and adjusting the diagonal elements according to the predicted \textit{Gaia} performance -- relative to DR3, parallax precisions in DR4 will improve by a factor of $\sim$1.33 and proper motion precisions by a factor of  $\sim$2.4\footnote[8]{\url{https://www.cosmos.esa.int/web/gaia/science-performance}, see also \citet{Brown2019}.} Off-diagonal elements of the covariance matrix remain unchanged.}

\textcolor{black}{We estimate \textit{Gaia} DR2/EDR3, DR3 and DR4 radial velocity errors for each star based on its Johnson-Cousins $V$-band magnitude and spectral type using the \texttt{PYTHON} package \texttt{PyGaia}\footnote[9]{\url{https://github.com/agabrown/PyGaia}}. Relative to DR2/EDR3, radial velocity errors in DR3 and DR4 improve by a factor of $\sim$1.33 and $\sim$1.65, respectively. We assume for all stars that the radial radial velocity uncertainties are uncorrelated to the astrometric uncertainties.}

\subsection{Identifying \textcolor{black}{high-confidence} HVSs}

\textcolor{black}{An ejected star} appearing in a \textit{Gaia} data release does not necessarily mean it will easily stand out as a promising HVS candidate. \textcolor{black}{Stars ejected} from the GC at low velocities will be difficult to distinguish from more typical Milky Way stars moving at similar speeds, and an unbound HVS cannot be identified as such if the \textcolor{black}{observational errors on its position and velocity are large} -- its true Galactocentric total velocity will be uncertain and its origin will be ambiguous.

We select which of our mock \textcolor{black}{ejected stars} would be conspicuous as promising HVS candidates taking inspiration in particular from existing HVS searches in the radial velocity catalogues of \textit{Gaia} DR2 \citep{Marchetti2019} and EDR3 \citep{Marchetti2021}. Following these works, We first select only mock stars with high-precision parallax uncertainties $(\sigma_\varpi/\varpi<20\%)$, for whom estimating a distance by inverting the parallax is non-problematic \citep{BailerJones2015}. \textcolor{black}{For each remaining star we sample one thousand times over the mock \textit{Gaia} astrometric and radial velocity uncertainties (see Sec. \ref{sec:methods:errors})}. From these, we compute the star's total velocity in the Galactocentric rest frame. We then compute $P_{\rm ub}$, the fraction of realizations in which the star is moving in excess of the Galactic escape speed at its position. We calculate escape velocities assuming the best-fit Galactic potential of \citet{McMillan2017}. We then select only stars which satisfy $P_{\rm ub}>0.8$, i.e. those which would stand out as being likely unbound to the Galaxy. Our results are not impacted if we compute escape velocities assuming other potentials commonly used in the context of HVSs, e.g. those outlined in \citet{Irrgang2013}, \citet{Bovy2015} or \citet{Rossi2017}. 

To reiterate and summarize, our clean sample of mock ejected stars which would appear as promising HVS candidates in the \textit{Gaia} DR2, DR3 and DR4 radial velocity catalogues consists of stars which have the following properties:
\begin{itemize}
    \item An apparent magnitude which satisfies the spectroscopic selection function of the corresponding data release.
    \item An effective temperature in the appropriate range for the given data release to be assigned a validated radial velocity.
    \item A relative parallax uncertainty less than 20\%.
    \item A probability 80\% or greater of being identified as unbound to the Galaxy according to its mock \textit{Gaia} astrometric uncertainties.
\end{itemize}

In the interest of brevity, for the remainder of this work when we use the terms \textit{Gaia} DR2/DR3/DR4 we are referring exclusively to the radial velocity subsets thereof, and by the term HVS we refer only to those stars which satisfy the criteria above. We also note that we opt not to discuss \textit{Gaia} EDR3 predictions in particular. In practice, we predict an identical number of HVSs in EDR3 as in DR2 \textcolor{black}{since the spectroscopic selection function remains unchanged} -- the improved astrometric precision does not result in significantly more HVSs, as the strict effective temperature and $G_{\rm RVS}$ magnitude criteria are the limiting factors.

\section{Existing Galactic Centre Constraints} \label{sec:constraints}

\textcolor{black}{In Secs. \ref{sec:methods:ejection} and \ref{sec:methods:tflight} we described our assumptions for the NSC IMF, the orbital period and mas distributions among NSC binaries, and the GC HVS ejection rate. These are summarized in Table \ref{tab:vars}. Here we outline how these assumptions are based on known literature constraints from observations or theoretical modeling. We additionally justify the ranges of $\kappa$, $\gamma$, $\pi$  and $\eta$ we explore in our model variations.} 

There have been several independent lines of evidence to suggest that that IMF in the YSC is top-heavy when compared to typically used canonical Galactic IMFs. \citet{Nayakshin2005} compared \textit{Chandra} X-ray observations of the GC region to the Orion nebula. They found that X-ray-bright low-mass ($m\leq3 \, \mathrm{M_\odot}$) stars in the GC are suppressed by a factor of at least ten compared to expectations assuming a \citet{Miller1979} IMF. Near-infrared observations using the SINFONI integral field spectrograph on the European South Observatory's Very Large Telescope have been instrumental in furthering our understanding of the GC. The K-band luminosity function of both early-type \citep{Paumgard2006} and late-type \citep{Maness2007}  SINFONI-identified stars within $1 \, \mathrm{pc}$ of the GC are consistent with a power-law initial mass function $f(m)\propto m^{\kappa}$ with $-1.35<\kappa<-0.85$. An even steeper IMF is suggested for SINFONI-identified YSC stars by \citet[][hereafter \citetalias{Bartko2010}]{Bartko2010} who find $\kappa=-0.45\pm0.3$. Note \citet{Lockmann2010}, however, who compare mass-to-light ratios of synthetic stellar population models to the GC and find that a deviation from a \citet[][hereafter \citetalias{Salpeter1955}]{Salpeter1955} IMF ($\kappa=-2.35$) is not required in the GC, though mass functions flatter than $\kappa\simeq-1$ cannot be ruled out. \textcolor{black}{Our fiducial $\kappa=-1.7$ is based on \citet[][hereafter \citetalias{Lu2013}]{Lu2013}} who use Keck near-infrared observations of young stars within $0.5 \, \mathrm{pc}$ of the GC to derive a top-heavy IMF with $\kappa=-1.7\pm0.2$. In this work when we vary our models we will explore $-0.3\leq\kappa\leq-3.3$ to capture the full range of top-heavy IMFs as well as those up to a $\sim$dex more steep than a standard \citetalias{Salpeter1955} IMF.

Studying the intrinsic properties of early-type GC binaries is particularly difficult, as the dust-obscuration and source-crowding hurdles of observing the GC in general are compounded with the difficulty of detecting even intermediate-mass companions around massive stars \citep[see][]{Kobulnicky2007,Duchene2013}. To date, only three O/Wolf-Rayet binaries have been detected within $0.2 \, \mathrm{pc}$ of Sgr A* \citep{Ott1999,Martins2006,Pfuhl2014}. To gain an intuition for the properties of GC binaries, we turn to the statistics of O/B binary populations elsewhere in the Milky Way and the Magellanic Clouds, particularly in environments rich in massive stellar populations such as Cygnus OB2 \citep{Kobulnicky2007, Kiminki2012, Kobulnicky2014}, SCo OB2 \citep{Kouwenhoven2007, Rizzuto2013} and 30 Doradus \citep{Sana2013, Dunstall2015, Almeida2017}. These studies generally find that the orbital log-periods ($\mathrm{log_{10}}P$) and mass ratios ($q$) of O/B binaries are consistent with power laws (i.e. $f(\mathrm{log_{10}}P)\propto (\mathrm{log_{10}}P)^{\pi}$, $f(q)\propto q^{\gamma}$), where $\pi$ is typically $\sim$0 or slightly negative and $\gamma$ is in the range [-1,0] \citep[e.g.][]{Kobulnicky2007, Sana2012, Sana2013, Moe2013, Kobulnicky2014, Moe2015, Banyard2021}. Note \citet{Dunstall2015}, however, who find a very steeply decreasing $q$ distribution ($\gamma=-2.8\pm0.8$) among B-type binaries in the VLT-FLAMES survey of 30 Doradus. In our model variations we explore $\pi$ in the range [-2,+2] and $\gamma$ in the range [-3,+2] to cover the full range of reported values in the literature. We do not consider correlations among $m_{\rm p}$, $\mathrm{log_{10}}P$ and $q$ as proposed by e.g. \citet{Moe2017} and \citet{Tokovinin2020}.

Finally, we consider the ejection rate $\eta$ of HVSs from the GC, a quantity which remains relatively unconstrained. Theoretical analytical estimates are typically in the range $10^{-5} \, \mathrm{yr^{-1}} \leq \eta \leq 10^{-3}  \, \mathrm{yr^{-1}}$ \citep{Hills1988,Yu2003} depending on how quickly stars are scattered into `loss cone' -- the phase space of close-periapse, low-angular momentum orbits about Sgr A*. Detailed simulations by \citet{Zhang2013} of GC dynamics favour an ejection rate of $10^{-5} \, \mathrm{yr^{-1}} \leq \eta \leq 10^{-4}\, \mathrm{yr^{-1}}$. A similar rate is obtained when one calibrates to the population of known HVS candidates \citep{Bromley2012, Brown2014, Marchetti2018}, and when one infers an HVS ejection rate from the rate of tidal disruption events in the nearby Universe \citep[see][]{Bromley2012, Brown2015rev}. In our model variations we consider rates in the range $10^{-6} \, \mathrm{yr^{-1}} \leq \eta \leq 10^{-1}  \, \mathrm{yr^{-1}}$ to capture the extrema of the plausible values.

\section{Results} \label{sec:results}

In this Section, we outline the number of \textcolor{black}{high-confidence} HVSs we expect to be present in \textit{Gaia} DR2, DR3 and DR4. We explore how these predictions depend on the model parameters $\kappa$, $\eta$, $\pi$ and $\gamma$ (see Sec. \ref{sec:constraints}) and show that the numbers of HVSs detected in DR3 and DR4 in the future will offer constraints on these parameters, even if those numbers are zero.

\subsection{Predicting the \textit{Gaia} HVS population}

To gain an intuition for the number $N_{\rm HVS}$ of HVSs we should expect to lurk in  \textit{Gaia} DR2/DR3/DR4, we first explore the relationship between $N_{\rm HVS}$ and the HVS ejection rate $\eta$. Since the ejection rate is constant, $N_{\rm HVS}$ will increase linearly with $\eta$. Sampling $\kappa$, $\pi$ and $\gamma$ uniformly in the ranges listed in Table \ref{tab:vars}, we eject and propagate 2000 HVSs samples. Later on in this section, we will assess the impacts of these parameters individually. 

\begin{figure}
    \centering
    \includegraphics[width=\columnwidth]{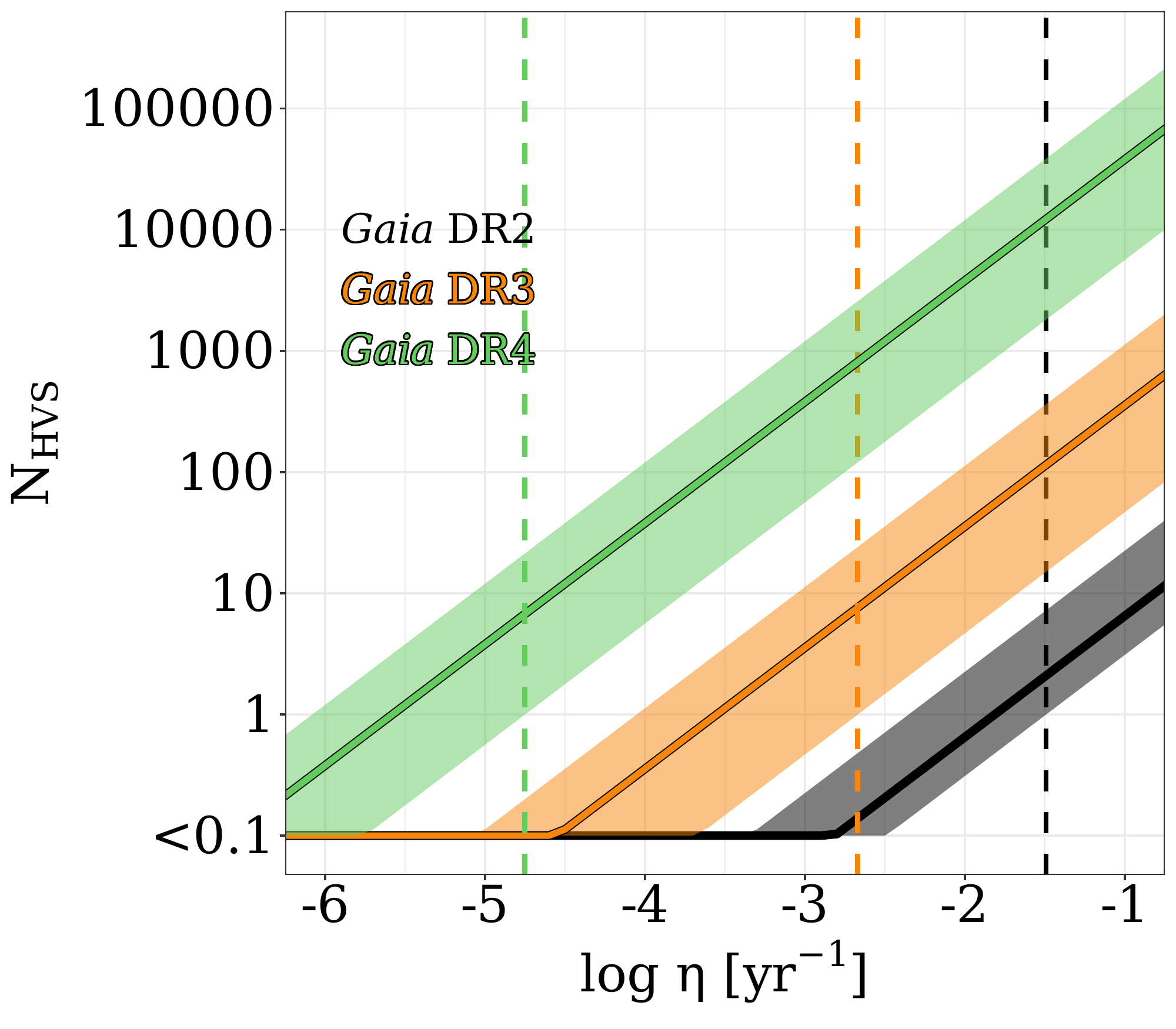}
    \caption{The number $N_{\rm HVS}$ of HVSs predicted to appear in the radial velocity catalogues of \textit{Gaia} DR2 (black), DR3 (orange) and DR4 (green) against the HVS ejection rate from the GC $\eta$, assuming an ejection model with flat priors on the initial mass function power law index $\kappa$, the binary log-period power law index $\pi$ and the binary mass ratio power law index $\gamma$. Models which predict $N\leq0.1$ are collapsed together. Shaded regions span the 1$\sigma$ confidence intervals computed over 2000 iterations. Vertical dashed lines show the ejection rate at which the 1$\sigma$ lower limit of $N_{\rm HVS}$ reaches 1 in the corresponding data release.}
    \label{fig:Nvseta}
\end{figure}

We show our results in Fig. \ref{fig:Nvseta}. The solid lines show how the expectation value for $N_{\rm HVS}$ changes with $\eta$ for \textit{Gaia} DR2, DR3 and DR4. The shaded regions show the 1$\sigma$ scatter of $N_{\rm HVS}$ over 2000 iterations. The vertical dashed lines show the ejection rates at which the lower limit of $N_{\rm HVS}$ reaches unity, i.e. the ejection rates at which one HVS \textit{at minimum} is predicted to appear in each data release. We see that <1 HVS is expected to exist in \textit{Gaia} DR2 for all but the most optimistic assumptions for $\eta$. At  our fiducial ejection rate of $\eta=10^{-4} \, \mathrm{yr^{-1}}$, there is only a $\lesssim$1\% chance of a high-confidence HVS appearing in DR2, consistent with previous \textcolor{black}{searches for fast stars} in this data release \citep{Hattori2018, Marchetti2019}. The prognosis for DR3 is slightly more optimistic; the expectation value for $N_{\rm HVS}$ is $\geq1$ as long as $\eta \gtrsim 3\times10^{-4} \, \mathrm{yr}^{-1}$ and one HVS at minimum should lurk in the survey if $\eta\gtrsim2\times10^{-3} \, \mathrm{yr^{-1}}$. Yet more HVSs are expected in DR4; at least one HVS is expected as long as $\eta\gtrsim10^{-5} \, \mathrm{yr^{-1}}$. This number rises to potentially $\sim$tens of thousands for the highest ejection rate assumptions.

\begin{figure*}
    \centering
    \includegraphics[width=2\columnwidth]{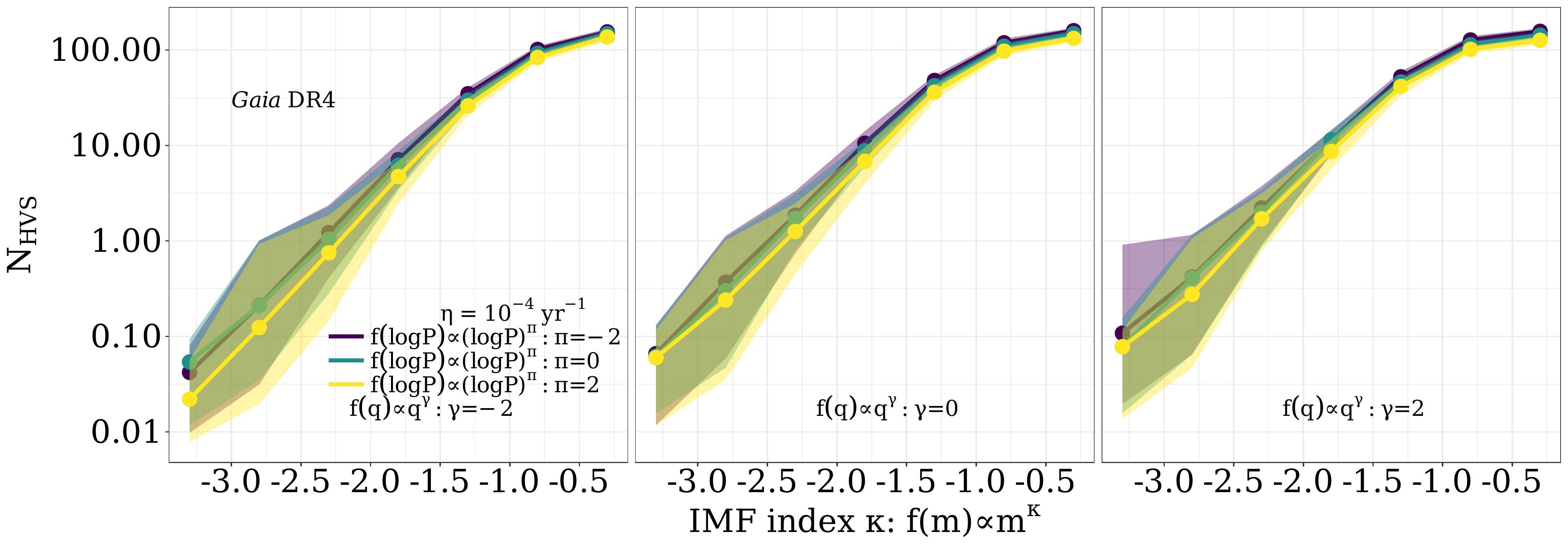}
    \includegraphics[width=2\columnwidth]{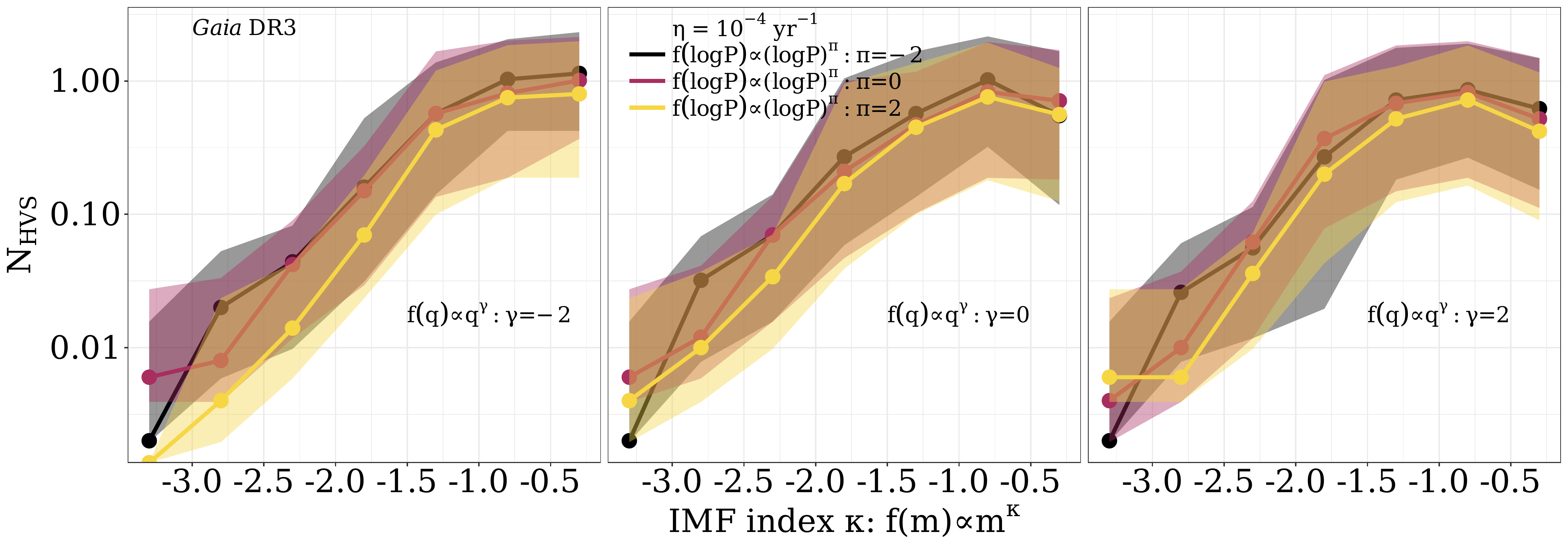}
    \caption{The number $N_{\rm HVS}$ of detectable high-confidence HVSs in the radial velocity catalogues of \textit{Gaia} DR4 (top row) and DR3 (bottom row) against the IMF power law index $\kappa$. Coloured lines show $N_{\rm HVS}$ for different binary log-period power law indices $\pi$. Panels show different binary mass ratio power law indices $\gamma$. An ejection rate of $\eta=10^{-4} \, \mathrm{yr^{-1}}$ is assumed for all models. Shaded regions span the 16th to 84th percentiles over 100 iterations.}
    \label{fig:NHVSkappa}
\end{figure*}

We next explore how the number of \textit{Gaia} HVSs changes with the NSC initial mass function slope $\kappa$ and the NSC binary log-period and mass ratio distribution power law slopes $\pi$ and $\gamma$. Given that <1 HVS is expected in \textit{Gaia} DR2 for all but the largest ejection rate assumptions, we focus on \textit{Gaia} DR3 and DR4. With the HVS ejection rate fixed at our fiducial value of $\eta=10^{-4} \, \mathrm{yr^{-1}}$, we show these predictions for \textit{Gaia} DR4 in the top row of Fig. \ref{fig:NHVSkappa}. We plot \textcolor{black}{the expectation and 1$\sigma$ scatter of $N_{\rm HVS}$ over 500 iterations  against $\kappa$ when $\pi$ and $\gamma$ are fixed.} The different lines show results for different $\pi$ and different panels for different $\gamma$. Most \textcolor{black}{striking} is the strong dependence of $N_{\rm HVS}$ on $\kappa$ -- increasing $\kappa$ from -3 to -1 increases the number of expected HVSs by three orders of magnitude. Towards the high end of the $\kappa$ range we explore, $N_{\rm HVS}$ seems to plateau, likely due to the fact that in such top-heavy IMFs the ejected HVSs will have quite short main sequence lifetimes. \textcolor{black}{$N_{\rm HVS}$ decreases slightly with increasing $\pi$ and with decreasing $\gamma$, however, these dependencies are far subdominant to the dependence of $N_{\rm HVS}$ on $\kappa$ and $\eta$. The scatter of $N_{\rm HVS}$ is due to the inherently stochastic nature of our simulations and the fact that the HVS population is propagated through a slightly different gravitational potential in each iteration (Sec. \ref{sec:methods:potential}).} 

In the second row of Fig. \ref{fig:NHVSkappa} we show predictions for \textit{Gaia} DR3. \textcolor{black}{We see similar results as for DR4, albeit with more scatter.} $N_{\rm HVS}$ depends most strongly on $\kappa$ and only weakly on $\pi$ and $\gamma$. Unlike our DR4 results, however, there is a hint of a turnover beyond $\kappa\gtrsim-1$, again due to the fact that an increasing fraction of ejected HVSs leave the main sequence before being observed. Note that assuming $\eta=10^{-4} \, \mathrm{yr^{-1}}$, less than one HVS is expected in \textit{Gaia} DR3 unless the NSC IMF slope is quite top-heavy.

\textcolor{black}{Overall, the dependence of $N_{\rm HVS}$ on $\kappa$, $\pi$ and $\gamma$ in both panels of Fig. \ref{fig:NHVSkappa} follows from Fig. \ref{fig:vejs} -- $N_{\rm HVS}$ increases when more $m\geq0.5 \, \mathrm{M_\odot}$ stars are ejected at large velocities, i.e. when the IMF is shallow (less negative $\kappa$) and when short orbital periods (more negative $\pi$) and mass ratios closer to unity (more positive $\gamma$) are favoured among the HVS progenitor binaries. Of these, the IMF index has by far the largest impact. Considering the more subtle impact of $\pi$ and $\gamma$ and the few HVSs and significant $N_{\rm HVS}$ scatter expected in \textit{Gaia} DR3 and DR4 for all but the largest ejection rates/shallowest IMF slopes, it will be difficult to offer meaningful constraints on the distributions of orbital periods and mass ratios among GC binaries using \textit{Gaia} DR3 and DR4 HVS counts alone.} For the remainder of this work we therefore consider only the constraints $N_{\rm HVS}$ can offer on the GC IMF slope and the HVS ejection rate. We perform a separate suite of simulations where we finely explore a grid of models in $\kappa-\eta$ space and sample $\pi$ and $\gamma$ in the ranges listed in Table \ref{tab:vars} \textcolor{black}{with uniform probability}. For each $\kappa-\eta$ grid point we eject and propagate 1000 \textcolor{black}{populations of stars ejected via the Hills mechanism}. 

\begin{figure*}
    \centering
    \includegraphics[width=\columnwidth]{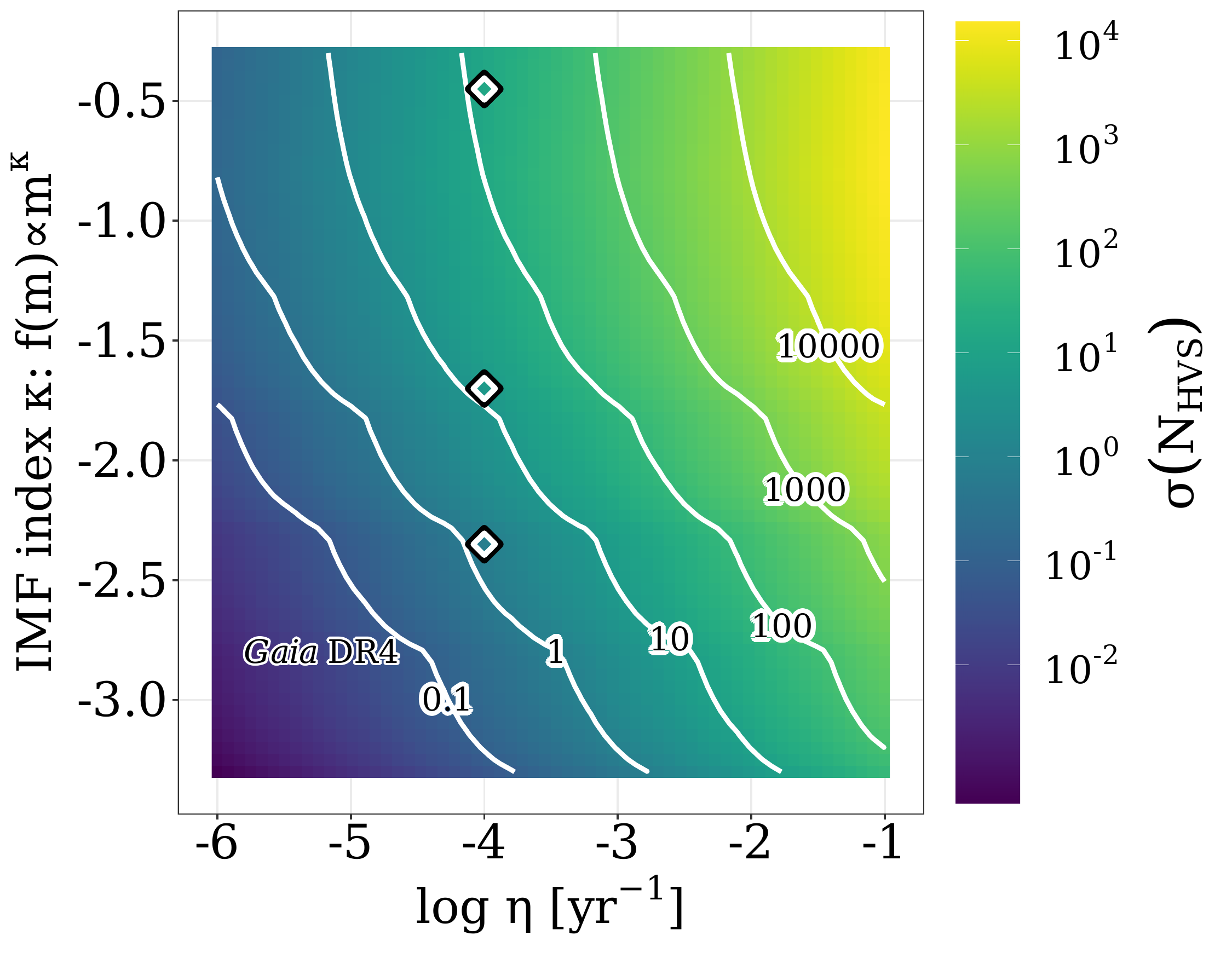}
    \includegraphics[width=\columnwidth]{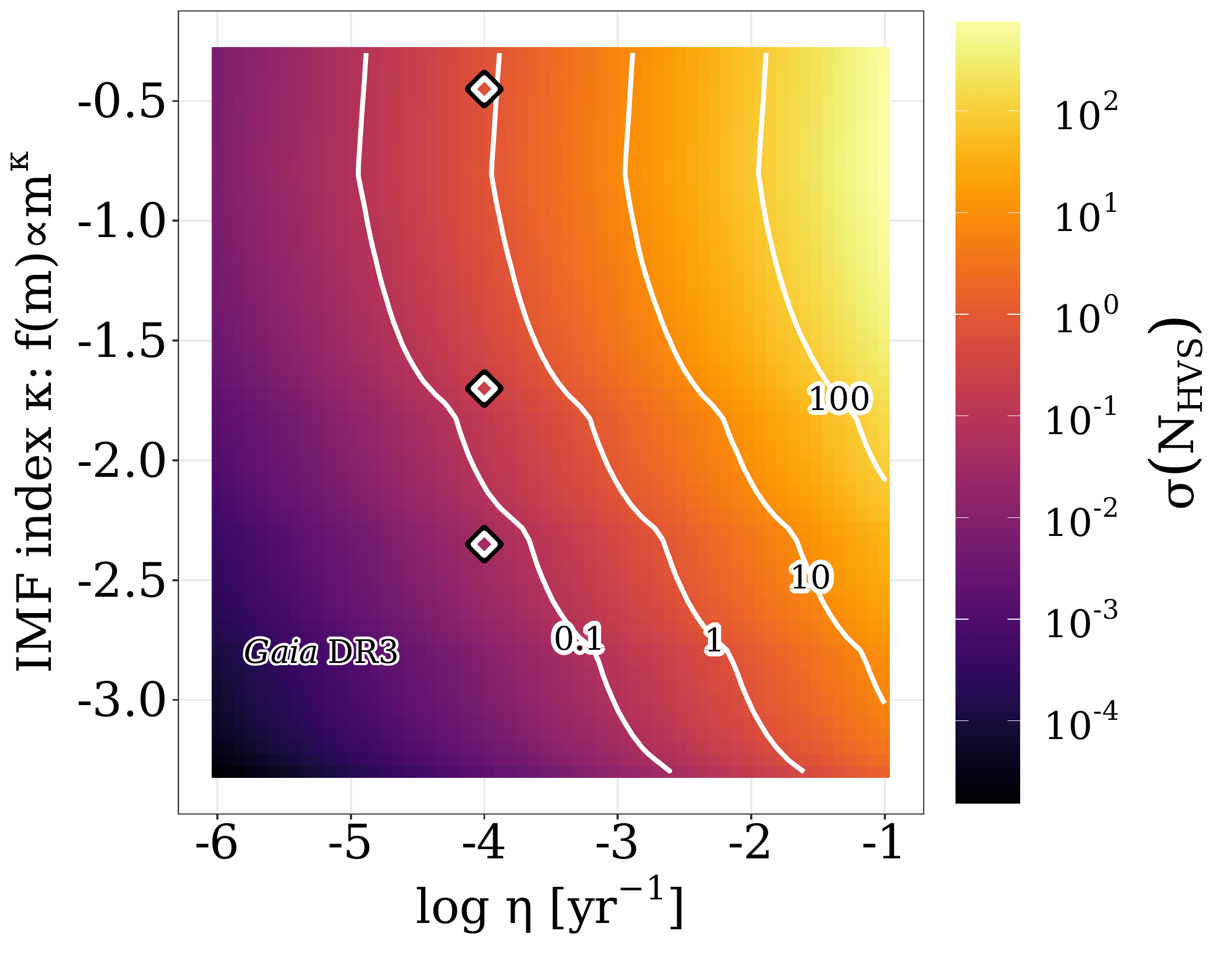}
    \caption{Contour lines show how $N_{\rm HVS}$ changes in the 2D parameter space of the IMF power law index $\kappa$ and the HVS ejection rate $\eta$ for both \textit{Gaia} DR4 (left) and DR3 (right), averaged over 1000 realizations. Colourbar shows how the 1$\sigma$ scatter of $N_{\rm HVS}$ evolves in this space. The black-and-white diamonds indicate our fiducial ejection rate of $\eta=10^{-4} \, \mathrm{yr^{-1}}$ and fiducial $\kappa=-1.7$ \citepalias{Lu2013}, as well as $\kappa=-2.35$ \citepalias{Salpeter1955} and $\kappa=-0.45$ \citepalias{Bartko2010}.}
    \label{fig:HVSkappaeta}
\end{figure*}

\textcolor{black}{In Fig. \ref{fig:HVSkappaeta} we show the evolution of $N_{\rm HVS}$ in the $\kappa-\eta$ space for \textit{Gaia} DR4 (left) and DR3 (right) with white contour lines.} The colourscales indicate the 1$\sigma$ scatter of $N_{\rm HVS}$ over the 1000 iterations, smoothed over the grid. The black-and-white diamonds indicate our fiducial $\kappa=-1.7$ \citepalias{Lu2013} and $\eta=10^{-4} \, \mathrm{yr^{-1}}$, as well as models with $\kappa$=-2.35 \citepalias{Salpeter1955} and $\kappa$=-0.45 \citepalias{Bartko2010} for comparison. We predict $13.8_{-4.8}^{+4.7}$, $1.3_{-0.4}^{+1.4}$ and $128\pm{13}$ HVSs to appear in \textit{Gaia} DR4 for each of these three models, respectively and $0.24_{-0.21}^{+0.78}$, $0.04_{-0.03}^{+0.02}$ and $0.8_{-0.7}^{+1.2}$ HVSs in DR3. So it is reasonable to expect at least one of these objects in \textit{Gaia} DR4, potentially hundreds or thousands for more optimistic ejection rates.

\subsection{On constraining the GC IMF and ejection rate}

\begin{figure}
    \centering
    \includegraphics[width =\columnwidth]{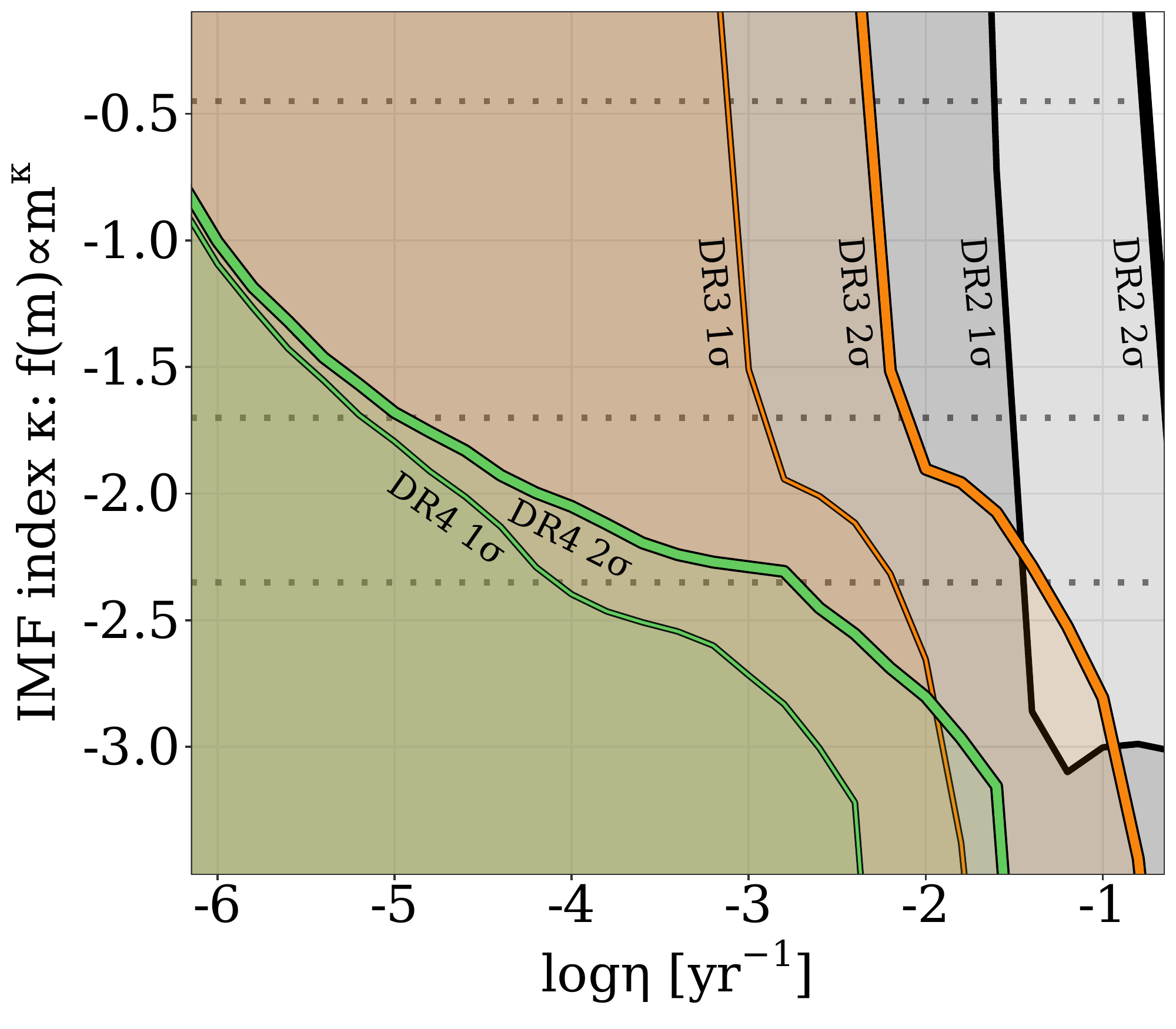}
    \caption{The regions below the thin (thick) coloured lines are areas in $\kappa-\eta$ parameter space are consistent within 1$\sigma$ (2$\sigma$) with finding zero \textit{Gaia} HVSs. Black, orange and green lines show constraints for \textit{Gaia} DR2, DR3, and DR4 respectively. Horizontal dotted lines indicate $\kappa=-2.35$ \citepalias{Salpeter1955}, $\kappa=-0.45$ \citepalias{Bartko2010} and $\kappa=-1.7$ \citepalias{Lu2013}.}
    \label{fig:HVSzero}
\end{figure}

Having investigated how the \textit{Gaia} DR2, DR3 and DR4 HVS populations depend on $\eta$, $\kappa$, $\pi$ and $\gamma$, and having established that $N_{\rm HVS}$ is most dependent on $\eta$ and $\kappa$, we can now answer more concrete questions: does the absence of high-quality GC-ejected HVSs in \textit{Gaia} DR2 \citep[see][]{Hattori2018, Marchetti2019} exclude certain combinations of $\kappa$ and $\eta$? If $N$ HVSs are discovered in \textit{Gaia} DR3 or DR4 in the future, how meaningfully will these constraints change?

We start with the most pessimistic case, in which the number of high-confidence HVSs detected in \textit{Gaia} DR2 remains at zero and in which despite high hopes, not a single promising HVS candidate is unearthed in \textit{Gaia} DR3 or DR4. The shaded regions in Fig. \ref{fig:HVSzero} show the combinations of $\kappa$ and $\eta$ for which the predicted 1$\sigma$ and 2$\sigma$ lower limits of $N_{\rm HVS}$ in \textit{Gaia} DR2/DR3/DR4 are less than 1. Any combination of $\kappa$ and $\eta$ within a shaded region is still allowed should zero HVSs be unearthed in the corresponding data release. The constraints provided by the current known lack of HVSs in \textit{Gaia} DR2 are not particularly strong -- only ejection rates above $2\times10^{-1} \, \mathrm{yr^{-1}}$ can be currently excluded with 2$\sigma$ confidence, and situations in which $\eta \gtrsim 3\times10^{-2} \, \mathrm{yr^{-1}}$ \textit{and} $\kappa\gtrsim-3$ can be currently excluded with 1$\sigma$ confidence. Regardless, these upper limits represent the first time \textit{Gaia} observations (or lack thereof) alone have been used to constrain the stellar population in the inner parsecs of our Galaxy.

Should zero HVSs be discovered in \textit{Gaia} DR3 and DR4, we will be able to place more discriminatory joint constraints on $\eta$ and $\kappa$. If future observations of the GC place tighter constraints on the NSC initial mass function then a complete lack of HVSs in \textit{Gaia} DR3 and especially in DR4 will place robust upper limits on the Galactic HVS ejection rate. The horizontal dotted lines show $\kappa=-2.35$ \citepalias{Salpeter1955}, $\kappa=-1.7$ \citepalias{Lu2013} and $\kappa=-0.45$ \citepalias{Bartko2010}. If zero HVSs are discovered in DR4, this will be inconsistent with predictions of a \citetalias{Bartko2010} IMF unless the HVS ejection rate is much lower than $10^{-6} \, \mathrm{yr^{-1}}$. If the GC IMF follows \citetalias{Lu2013} or \citetalias{Salpeter1955}, HVS ejection rates higher than $\sim10^{-5} \, \mathrm{yr^{-1}}$ and $\sim10^{-3} \, \mathrm{yr^{-1}}$ are excluded at 2$\sigma$ confidence, respectively. 

\begin{figure*}
    \centering
    \includegraphics[width =\columnwidth]{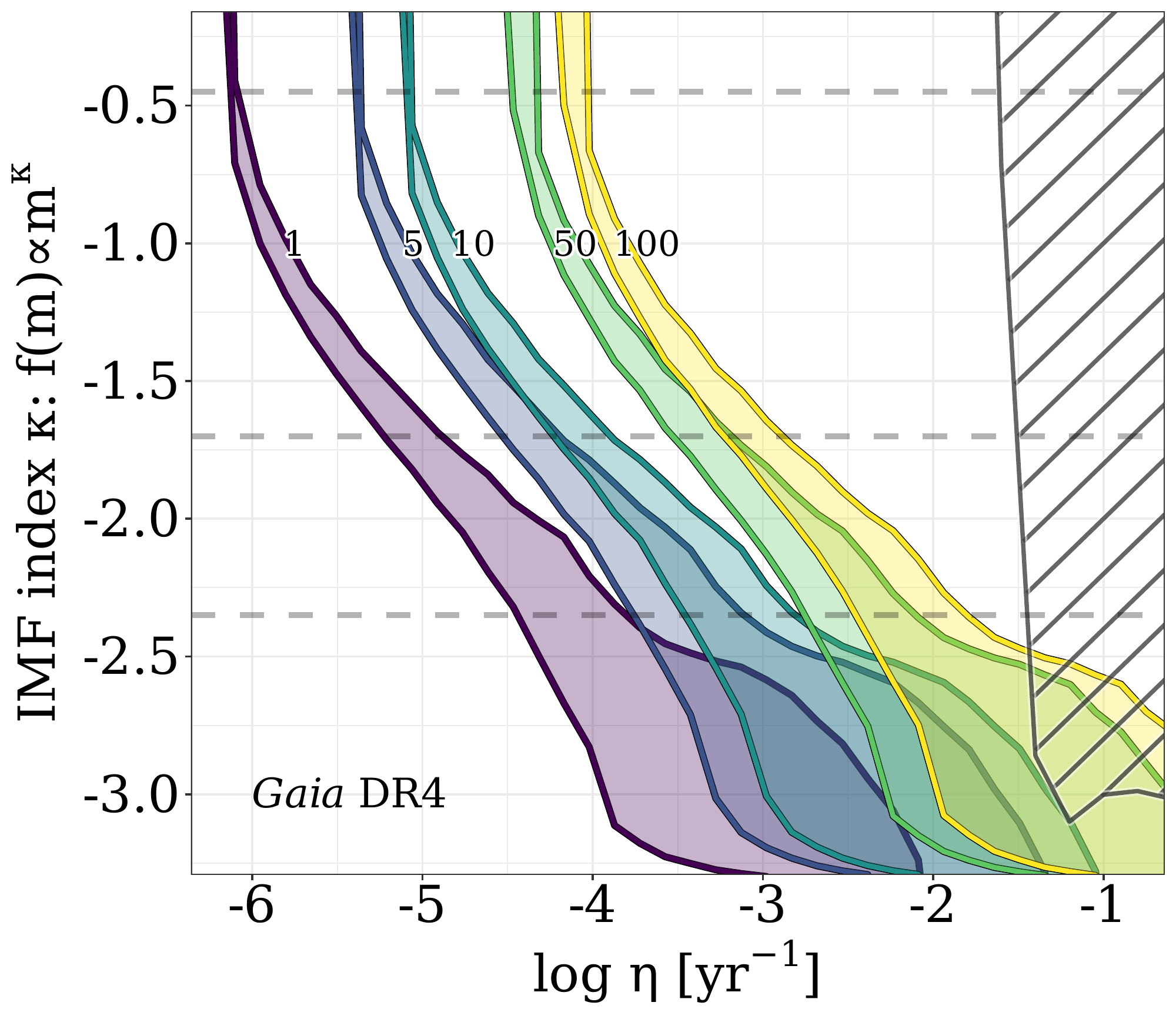}
    \includegraphics[width =\columnwidth]{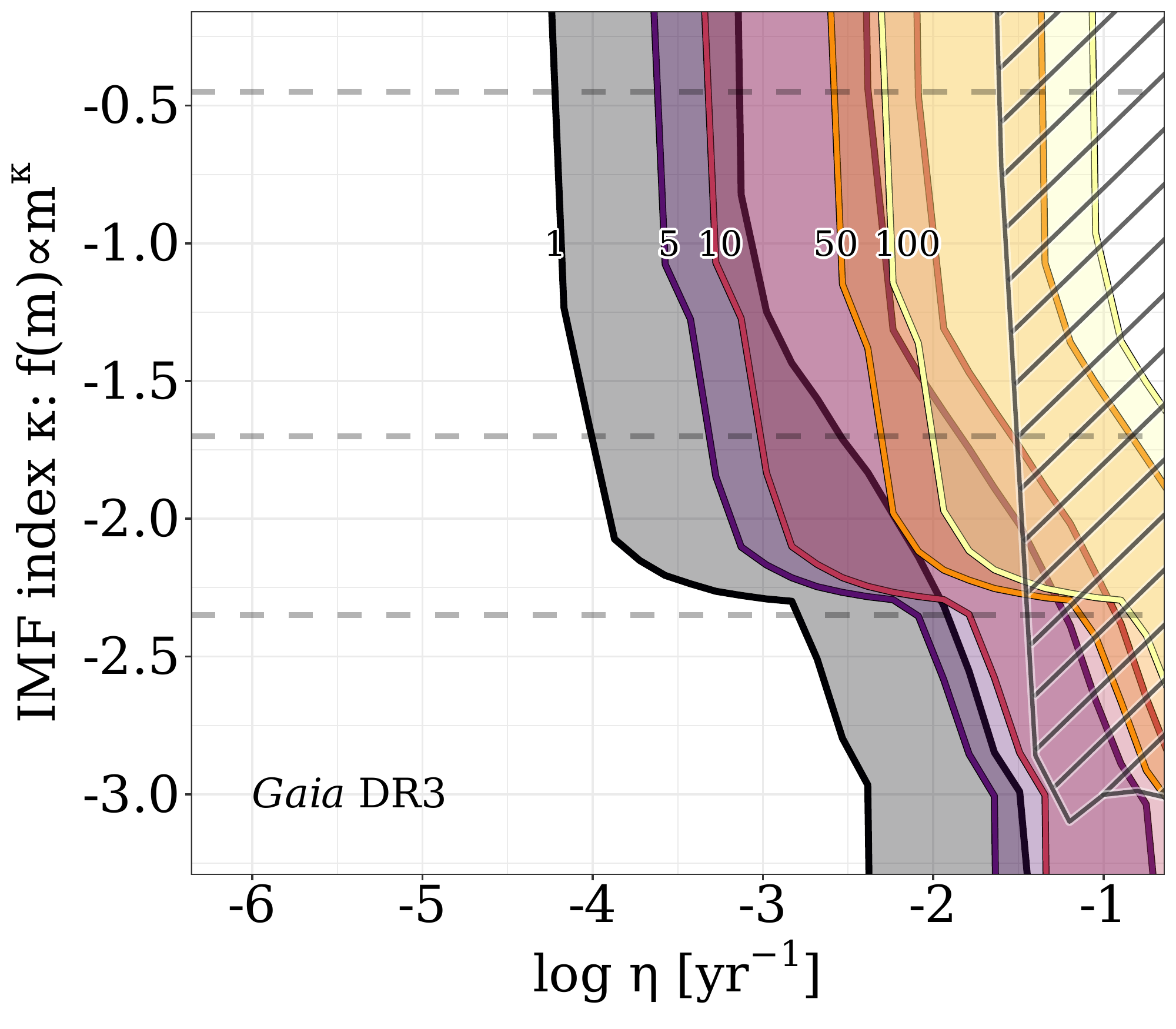}
    \caption{Similar to Fig. \ref{fig:HVSzero}, shaded coloured regions show the allowed regions of the $\kappa-\eta$ parameter space if 1/5/10/50/100 HVSs are discovered in \textit{Gaia} DR4 (left) and DR3 (right). Horizontal dashed lines indicate $\kappa=-2.35$ \citepalias{Salpeter1955}, $\kappa=-1.7$ \citepalias{Lu2013} and $\kappa=-0.45$ \citepalias{Bartko2010}. The hashed area indicates regions of $\kappa-\eta$ space currently excluded given the lack of detected HVSs in \textit{Gaia} DR2.} 
    \label{fig:HVSnumbers}
\end{figure*}

We next consider the case in which $N>0$ HVSs are discovered in \textit{Gaia} DR3 or DR4. The only allowed combinations of $\kappa$ and $\eta$ are those which predict a number of HVSs consistent with $N$. We illustrate this in Fig. \ref{fig:HVSnumbers}. Each coloured band in the left panel shows the region of $\kappa-\eta$ space consistent at the 1$\sigma$ level with discovering $N$ HVSs in \textit{Gaia} DR4. For instance, if a single high-confience HVS is discovered in DR4, the violet stripe shows the models in which the 1$\sigma$ range of the predicted HVS population size includes 1. No single IMF slope can be entirely excluded -- a steep IMF with a large ejection rate or a shallow IMF with a lower ejection rate can predict identical HVS population sizes. However, the joint constraints can be quite strict. For example, if 5 HVSs are discovered in DR4 and if the \citetalias{Lu2013} IMF slope $\kappa=-1.7\pm0.2$ is independently confirmed, the HVS ejection rate in the GC is restricted to the range $\eta\sim[2-20]\times 10^{-5} \, \mathrm{yr^{-1}}$. If 50 HVSs are discovered, the possible range becomes $\eta\sim[2-20]\times 10^{-4} \, \mathrm{yr^{-1}}$.

In the right panel of Fig. \ref{fig:HVSnumbers} we do the same analysis for \textit{Gaia} DR3. The bands of allowed regions are broader than for DR4 but can still provide meaningful constraints -- For a \citetalias{Lu2013} IMF, the ejection rate is restricted to $\eta\sim[1-30]\times 10^{-4} \, \mathrm{yr^{-1}}$ if one HVS is unearthed in DR3. Overall, a non-zero HVS population in DR3 implies an HVS ejection rate of $\eta\gtrsim10^{-4} \, \mathrm{yr^{-1}}$. A DR3 HVS population significantly in excess of 10 is only possible if the HVS ejection rate is $\eta\gtrsim 3\times10^{-3} \, \mathrm{yr^{-1}}$, or slightly lower if the NSC IMF is quite top-heavy.

 In both plots we show the region of $\kappa-\eta$ space excluded at 1$\sigma$ confidence by the lack of HVSs in \textit{Gaia} DR2 with the hashed area. Our results indicate that most configurations allowing for 100+ HVSs in \textit{Gaia} DR3 are already excluded, while configurations allowing for hundreds or thousands of HVSs in DR4 remain possible.

\begin{figure*}
    \centering
    \includegraphics[width=2\columnwidth]{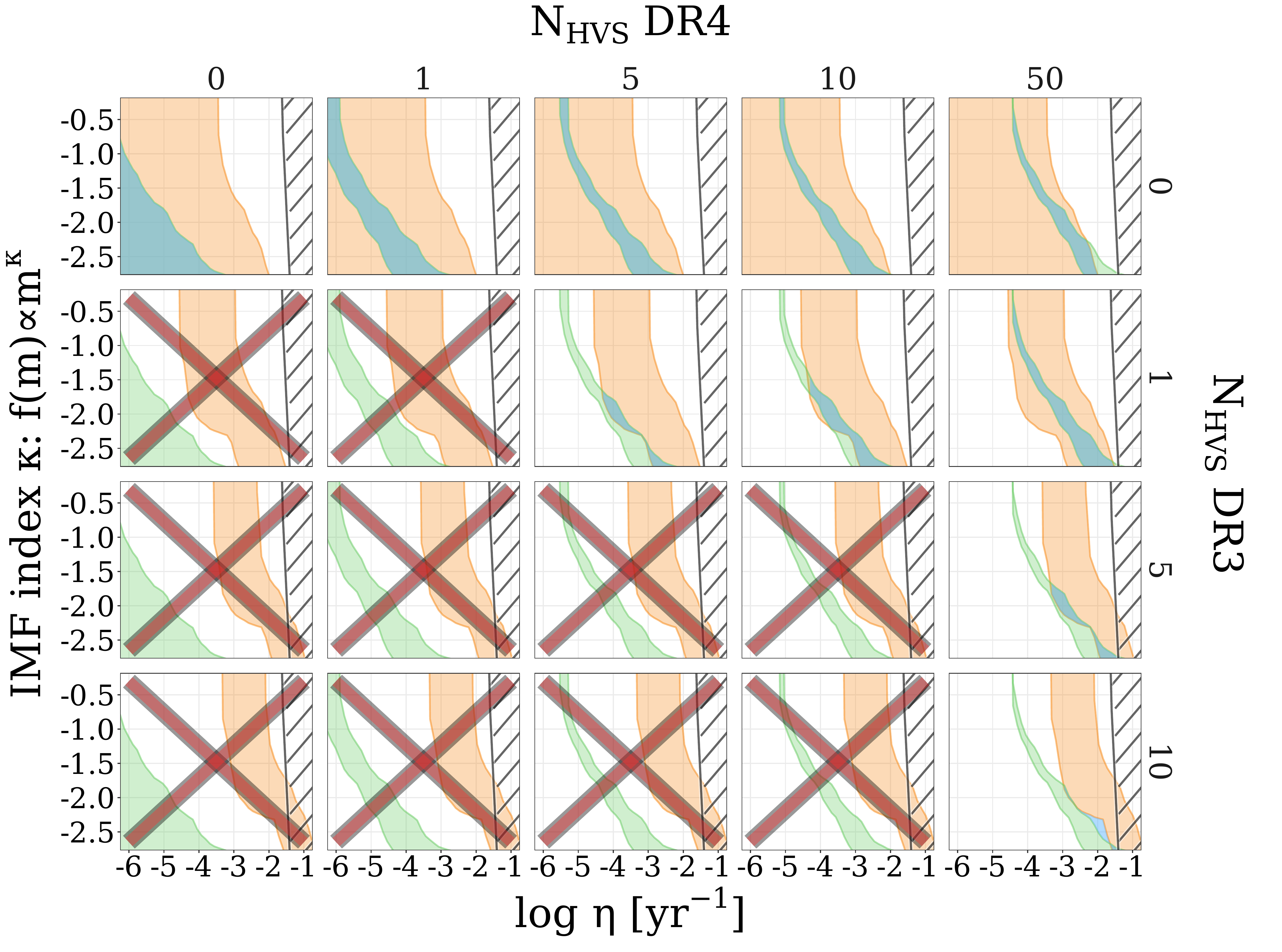}
    \caption{Similar to Fig. \ref{fig:HVSnumbers}, shaded orange and green regions respectively show the allowed regions of the $\kappa-\eta$ parameter space for a given number of uncovered HVSs in DR3 (rows) and DR4 (columns). Blue shaded regions show where the DR3 and DR4 constraints overlap. The hashed area indicates regions of $\kappa-\eta$ space currently excluded given the lack of detected HVSs in \textit{Gaia} DR2. Panels marked with an X have no overlapping DR3 and DR4 regions in this parameter space, i.e. they show combinations of DR3 and DR4 results which are not possible in our model.}
    \label{fig:N3N4}
\end{figure*}

Finally, we consider the joint constraints provided by the \textit{Gaia} DR3 and DR4 HVS populations. We have shown that when $N_{\rm DR4}\geq0$ HVSs are eventually discovered in DR4, it can place constraints on the GC IMF and HVS ejection rate (Figs. \ref{fig:HVSzero}, \ref{fig:HVSnumbers}). At this time, however, we will already know that DR3 contains $N_{\rm DR3}\geq0$ HVSs. Does this extra information meaningfully improve constraints? We explore this in Fig. \ref{fig:N3N4}. Similar to Fig. \ref{fig:HVSnumbers}, the orange and green bands in each panel show the individual allowed regions for combinations of $N_{\rm DR4}$ (columns) and $N_{\rm DR3}$ (rows). The overlap of these regions is highlighted in blue, i.e. $\kappa$ and $\eta$ \textit{must} lie in the blue region if $N_{\rm DR4}$ HVSs are in DR4 and $N_{\rm DR3}$ in DR3. For some $N_{\rm DR4}$/$N_{\rm DR3}$ combinations, e.g. for $N_{\rm DR3}$=0 and $N_{\rm DR4}$=1, the joint constraints are no different from the DR4 constraints alone. For other combinations, the joint constraints become quite restrictive -- if $N_{\rm DR3}$=5 and $N_{\rm DR4}$=50, for example, only a quite narrow slice of the $\kappa-\eta$ parameter space is allowed. Panels marked with a red X are impossible in our model. If ten HVSs are uncovered in DR3, it is not possible for DR4 to contain only five, as DR4 will presumably contain all ten DR3 HVSs, plus potentially more. 

\section{Discussion \& Conclusions}
\label{sec:discussion}

In this work we simulate the ejection of hyper-velocity stars (HVSs) from the Galactic Centre (GC) via the Hills mechanism. We predict the sizes of the HVS populations likely to lurk in future data releases from the European Space Agency's \textit{Gaia} satellite. We explore how these population sizes depend on the properties of the nuclear star cluster (NSC) in the GC, namely its stellar initial mass function (IMF), the mass ratio and orbital period distributions among its binaries, and the ejection rate of HVSs from the GC. We show that future detected HVS populations can place meaningful constraints on these properties. 

Throughout this work we have drawn star masses and ejection velocities assuming stars are ejected via the Hills mechanism \citep{Hills1988}. It is important to note that the \textit{Gaia} DR3 and DR4 constraints on $\kappa$ and $\eta$ that we forecast apply exclusively to a Hills ejection scenario. While the Hills mechanism remains the most popular proposed mechanism for ejecting HVSs, there are numerous alternative mechanisms which would still be effective at ejecting HVSs from the GC -- full suites of simulations exploring each mechanism in detail are outside the scope of this work. One alternative is a scenario in which Sgr A* has an as-of-yet undetected supermassive or intermediate-mass black hole companion. Single stars interacting with this arrangement can be ejected with velocities comparable to the Hills mechanism, though the ejection velocities and preferred ejection directions vary as the black hole binary spirals in \citep[e.g.][]{Yu2003,Gualandris2005, Baumgardt2006,Sesana2006, Sesana2007,Guillochon2015, Fragione2018, Darbha2019, Zheng2021}. Other potential ejection mechanisms in the GC include the interaction between single stars and a population of stellar-mass black holes in orbit around Sgr A* \citep{OLeary2008}, and the interaction between an infalling globular cluster with a single or binary massive black hole system at the GC \citep{Capuzzo2015, Fragione2016}.

In this study we make use of mock \textit{Gaia} astrometry, photometry and radial velocity measurements. While we account for uncertainties in these measurements, we implicitly assume zero systematic error. Of particular interest is the global parallax zero point, the mean offset between the observed and true parallax of a source. In \textit{Gaia} DR2 and EDR3 this offset is \textcolor{black}{estimated at} $-0.029 \, \mathrm{\mu as}$ \citep{Lindegren2018} and $-0.017 \, \mathrm{\mu as}$ \citep{Lindegren2021zp} respectively, but can vary by several tens of $\mathrm{\mu as}$ depending on the apparent magnitude, sky position, and colour of a \textit{Gaia} source. Being a negative offset, this means that distances inferred by inverting \textit{Gaia} parallaxes will be over-estimated on average. This means total velocities will be over-estimated as well once proper motions are converted to tangential velocities. Correcting for the zero point offset is known to reduce the number of objects identified as being likely unbound to the Galaxy in \textit{Gaia} DR2 and EDR3 \citep[][]{Marchetti2019, Marchetti2021}. Future HVS searches in both \textit{Gaia} and other surveys must continue to carefully consider the impacts of survey systematics, or else the constraints that HVSs can offer on the GC environment may be compromised. 

Our findings can be summarized as follows:

\begin{itemize}
    \item The numbers $N_{\rm HVS}$ of HVSs  gravitationally unbound to the Galaxy with high-precision astrometry and radial velocities in \textit{Gaia} Data Release 3 (DR3) and Data Release 4 (DR4) depend most strongly on the HVS ejection rate $\eta$ and the NSC IMF power law log-slope $\kappa$. $N_{\rm HVS}$ is comparatively less sensitive to the NSC binary orbital period distribution and to the NSC binary mass ratio distribution  (Figs. \ref{fig:Nvseta}, \ref{fig:NHVSkappa}).
    \item For a fiducial model in which $\kappa=-1.7$ \citep{Lu2013} and $\eta=10^{-4} \, \mathrm{yr^{-1}}$ \citep[see][]{Brown2015rev}, we predict $0.23_{-0.20}^{+0.79}$ HVSs in DR3 and  $12\pm5$ in DR4. If the NSC IMF is particularly top-heavy and the HVS ejection rate quite high, however, the \textit{Gaia} DR3 and DR4 populations could read several hundred and several tens of thousands, respectively (Fig. \ref{fig:HVSkappaeta}).
    \item The known lack of GC-ejected HVS candidates in \textit{Gaia} DR2 excludes models in which $\eta\gtrsim3\times10^{-2} \, \mathrm{yr^{-1}}$ \textit{and} $\kappa\gtrsim-3$. If zero HVSs are discovered in both \textit{Gaia} DR3 and DR4, the NSC IMF must be top-light and/or the HVS ejection rate must be low. For a model in which $\kappa=-1.7$ \citep{Lu2013}, the HVS ejection rate must be smaller than $\sim10^{-5} \, \mathrm{yr^{-1}}$ if zero HVSs are discovered in \textit{Gaia} DR4 (Fig. \ref{fig:HVSzero}).
    \item If $N_{\rm HVS}>0$ HVSs are unearthed in \textit{Gaia} DR3/DR4, tighter constraints can be placed on $\kappa$ and $\eta$. A degeneracy exists in this space, so meaningful constraints cannot be placed on these parameters individually by \textit{Gaia} HVS populations alone (Figs. \ref{fig:HVSnumbers}, \ref{fig:N3N4}).
\end{itemize}

We have shown in this work that with the upcoming \textit{Gaia} data releases, HVSs will begin to realise their full potential as an unorthodox yet effective tool for investigating the still-uncertain properties of the stellar populations at the centre of our Galaxy. The independent constraints they will offer on the initial mass function in the GC and the GC HVS ejection rate will synergise with existing constraints provided by direct observation and theoretical modeling, improving our understanding of Sgr A* and its impact on its environment.

\section*{Acknowledgements}

\textcolor{black}{We thank the anonymous referee for their feedback.} We thank Anthony Brown for helpful discussions, Andrew Everall and Douglas Boubert for helpful correspondence, and Paul McMillan for providing Milky Way potentials. FAE acknowledges funding support from the Natural Sciences and Engineering Research Council of Canada (NSERC) Postgraduate Scholarship. TM acknowledges an ESO fellowship. EMR acknowledges support from ERC Grant number: 101002511 / Project acronym: VEGA\_P.

\section*{Data Availability}

The simulation outputs underpinning this work can be shared upon reasonable request to the corresponding author.


\bibliographystyle{mnras}
\bibliography{HRS}
\appendix
\section{The properties of the \textit{Gaia} HVS population}

\begin{figure*}
    \centering
    \includegraphics[width=\columnwidth]{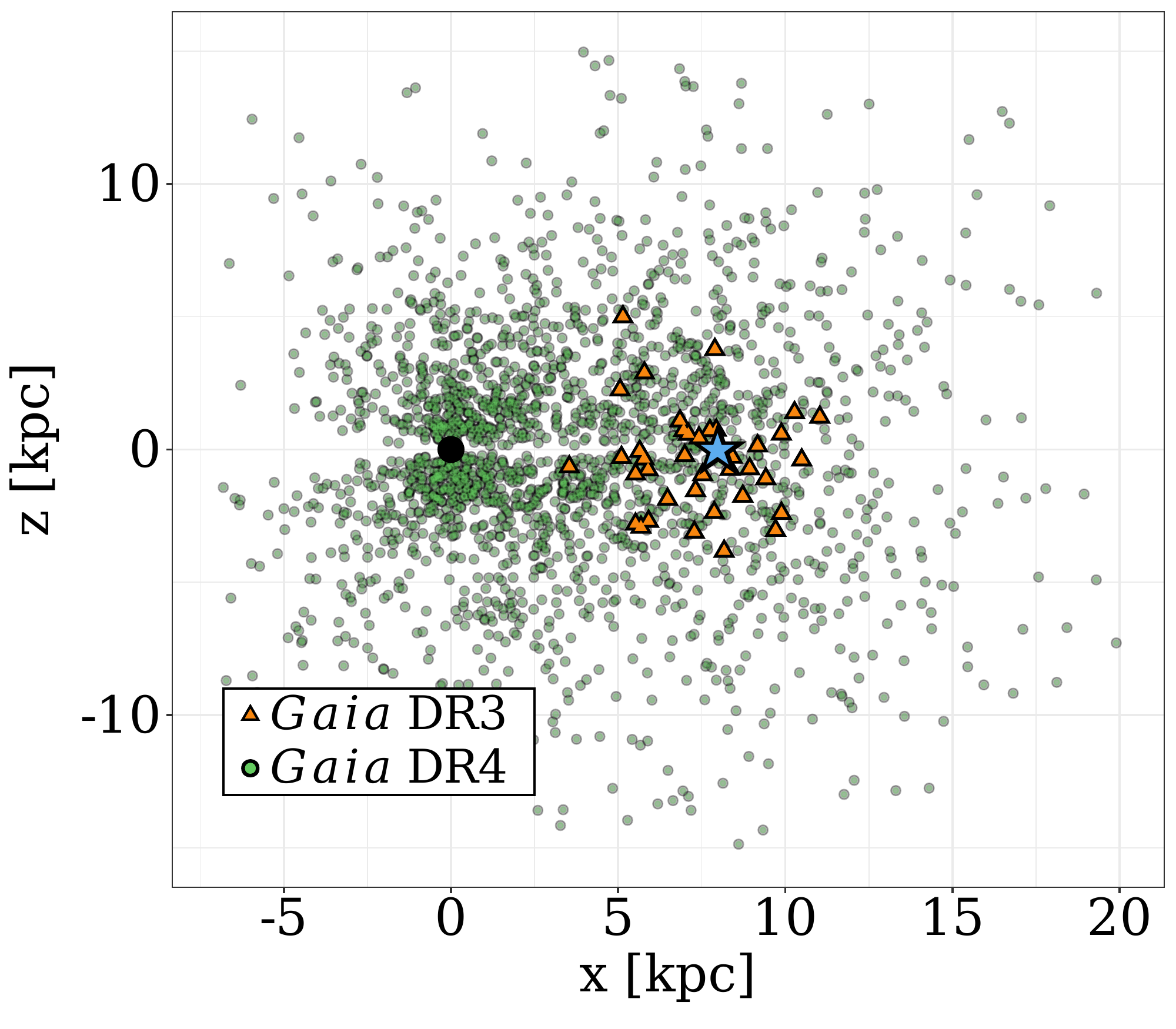} 
    \includegraphics[width=\columnwidth]{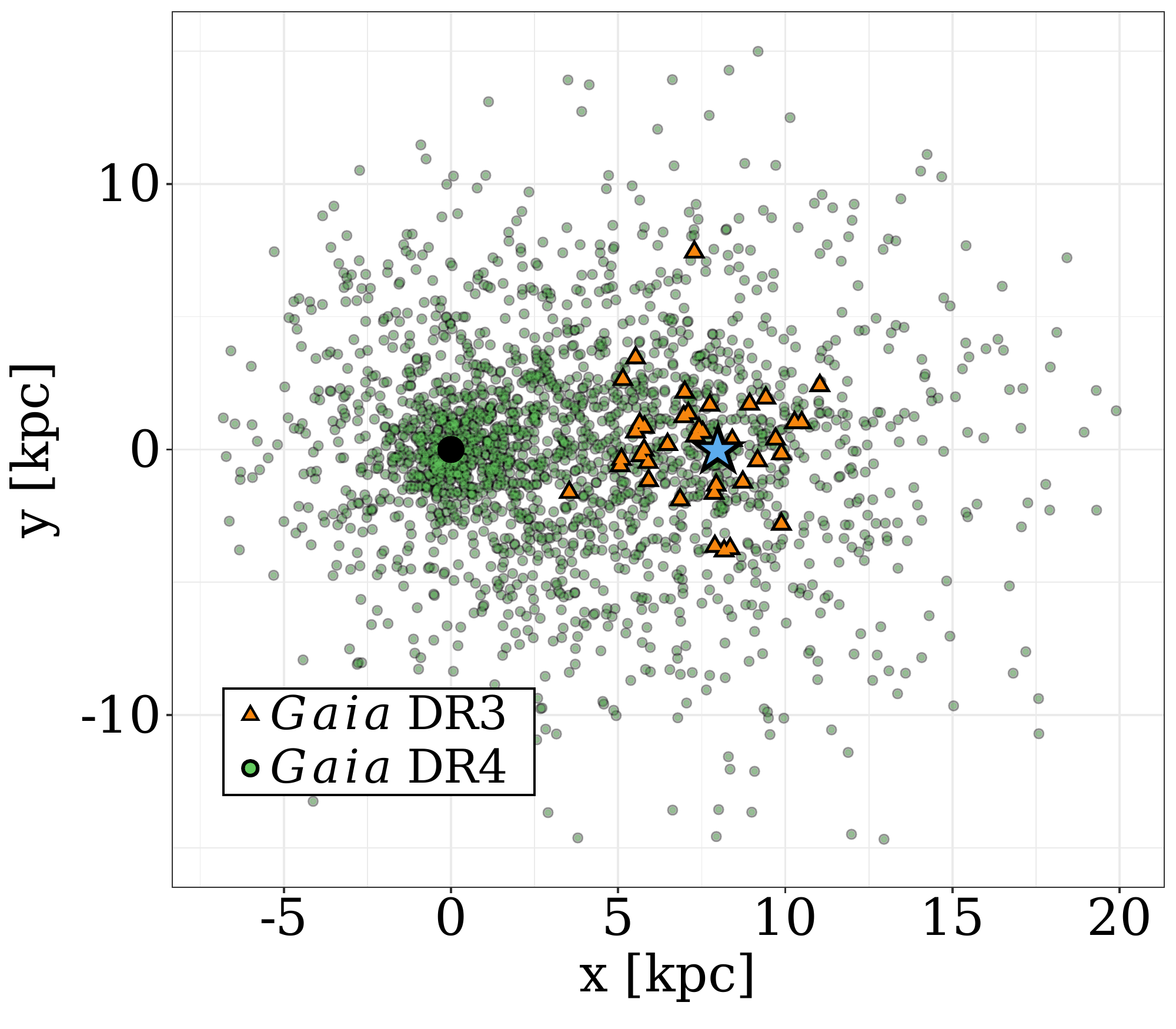} 
    \includegraphics[width=2\columnwidth]{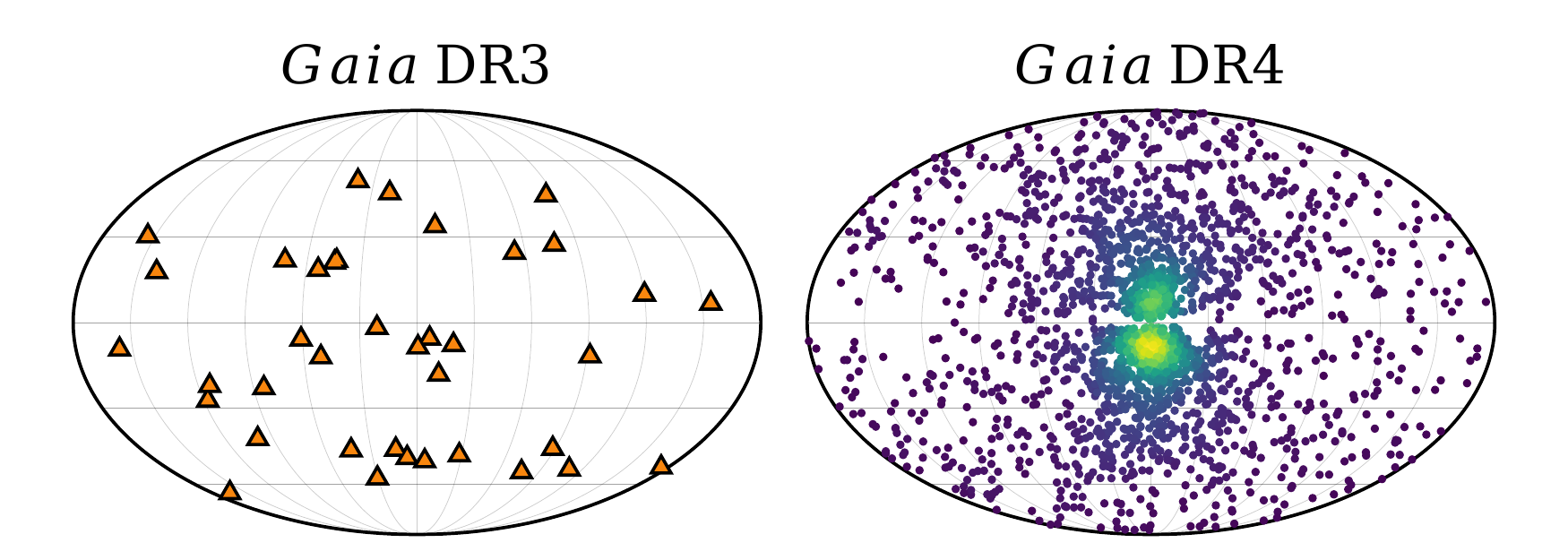} 
    \caption{\textit{Top row:} Distribution of HVSs in our fiducial model (see Sec. \ref{sec:constraints}) predicted appear in \textit{Gaia} DR3 (orange) and DR4 (green) in the Galactocentric Cartesian x-z (left) and x-y (right) planes. The position of the Sun at (x, y, z)$\simeq$(8, 0, 0)$\, \mathrm{kpc}$ is marked with a blue star. 200 stacked samples of HVSs are shown. \textit{Bottom row:} Distribution of the same predicted DR3 and DR4 HVS populations in a Mollweide projection in Galactic coordinates. Colour in the right panel indicates the density of points on a linear scale. 200 stacked samples of HVSs are shown.}
    \label{fig:Skypoints}
\end{figure*}

Figs. \ref{fig:Nvseta}, \ref{fig:NHVSkappa} and \ref{fig:HVSkappaeta} indicate that unless the GC IMF is particularly top-heavy or the ejection rate particularly high, the populations of HVSs in \textit{Gaia} DR4 and especially DR3 are not likely to be larger than a few tens. Obtaining useful summary statistics of their properties may therefore not be feasible. Nevertheless, it is interesting to ask where future \textit{Gaia} HVSs will be found and what properties we can expect of them.

In Fig. \ref{fig:Skypoints} we examine 200 stacked iterations of HVSs ejected from our fiducial model (see Table \ref{tab:vars}). In the top panels, we show how detectable HVSs populate the x-z (left) and x-y (right) plane of the Galaxy in Cartesian coordinates. The difference between the DR3 and DR4 magnitude limits is clear -- DR3 HVSs are confined to within $\sim7 \, \mathrm{kpc}$ of the Sun while DR4 HVSs are found up to $\sim 15 \, \mathrm{kpc}$ away. While strong dust attenuation prevents the detection of HVSs directly in the Galactic midplane, we see a clear concentration of HVSs near the GC in both planes -- a quarter of all DR4 HVSs will be found within $4 \, \mathrm{kpc}$ of the GC. In the bottom panels of Fig. \ref{fig:Skypoints} we show how DR3 (left) and DR4 (right) HVSs will populate the sky in Galactic coordinates. While the DR3 HVS(s) are distributed more or less uniformly across the sky, DR4 HVSs will \textcolor{black}{be found mostly at low latitudes near} the Galactic meridian. $30$\% of DR4 HVSs will be found within 25 degrees of $l=0^\circ,b=0^\circ$. The north-south asymmetry in Fig. \ref{fig:Skypoints} is due to the fact that the Galactic north is more dust-extincted than the south along $l\sim0^\circ$ \citep[see][]{Green2015, Bovy2015}.

\begin{figure*}
    \centering
    \includegraphics[width=2\columnwidth]{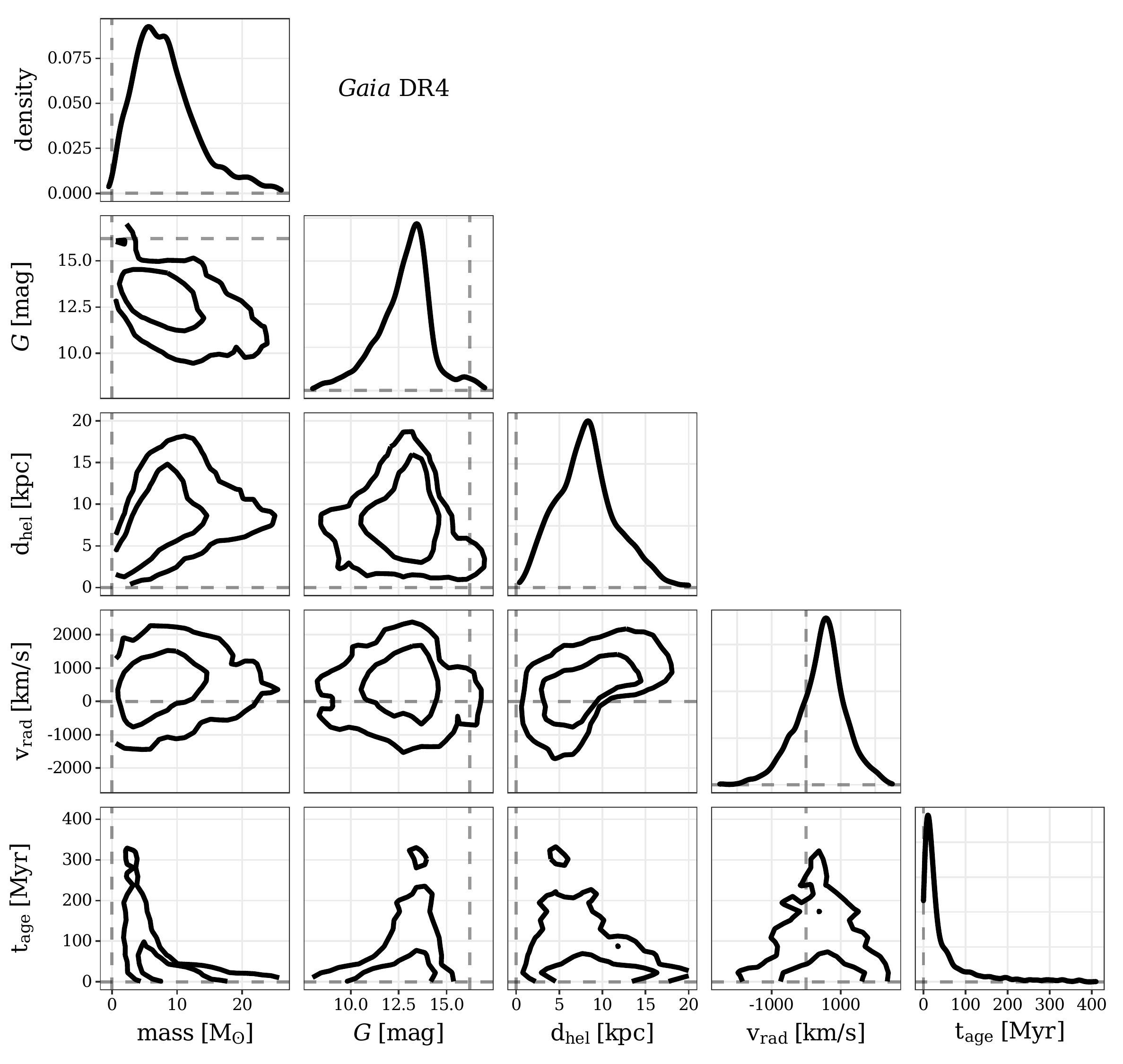} 
    \caption{The distributions of and between stellar mass $m$, \textit{Gaia G}-band magnitude, heliocentric distance $d_{\rm hel}$, radial velocity $v_{\rm rad}$ and stellar age $t_{\rm age}$ for HVSs detectable in \textit{Gaia} DR4. Contours show the 84th and 95th percentiles of the distributions. Distributions are calculated over 200 stacked samples of HVSs ejected from our fiducial model.}
    \label{fig:stairstep}
\end{figure*}

We summarize some kinematic and stellar properties of the DR4 HVSs ejected from our fiducial model in Fig. \ref{fig:stairstep}. We show distributions of and between stellar mass, apparent magnitude, heliocentric distance, radial velocity and stellar age. The prototypical \textit{Gaia} DR4 HVS is a $7.9_{-4.1}^{+6.3} \, \mathrm{M_\odot}$, $23_{-16}^{+107} \, \mathrm{Myr}$-old star $8.1_{-3.5}^{+3.7} \, \mathrm{kpc}$ away from the Sun. Notably,  the range of radial velocities is quite large -- while the most likely radial velocity of a DR4 HVS is $\sim$+500$ \, \mathrm{kpc \ s^{-1}}$, slow radial velocities and highly-negative radial velocities are possible as well depending on the flight time and ejection direction of the HVS. This can be somewhat counter-intuitive -- for the most part, currently-known HVS candidates are mostly massive stars in the Galactic halo on outbound trajectories, so this can give the impression that a highly positive radial velocity is a feature common to all HVSs.

\section{Contamination by unbound stars not ejected from the GC}
With the notable exception of HVS candidate S5-HVS1 \citep{Koposov2020}, it is notoriously difficult to indisputably associate known HVS candidates with an origin in the GC. For many candidates, tracing their trajectories backwards in time leads to a wide range of possible ejection locations \citep[see][]{Irrgang2019, Kreuzer2020, Irrgang2021} due to their imprecise astrometry and large heliocentric distances. Additionally, it is well-known that the GC is not the only potential birthplace of extreme-velocity stars. Many \textcolor{black}{known stars with estimated velocities above the Galactic escape speed} have past trajectories which imply an origin in the Galactic disc \citep[see][and references therein]{Evans2020}. The ejection rate of these so-called `hyper-runaway' stars is estimated to be a factor of 100 smaller than the ejection rate of HVSs \citep{Bromley2009, Kenyon2014}. Other ejection mechanisms must be invoked to explain them, such as dynamical ejections from dense stellar clusters \citep{Poveda1967, Leonard1990, Perets2012, Oh2016} and the ejection of companions from a tight binary following a core-collapse event \citep{Blaauw1961,Tauris1998, Portegies2000, Tauris2015, Renzo2019, Evans2020}. Others \textcolor{black}{unbound star candidates} have trajectories which suggest an origin outside the Milky Way \citep{Marchetti2019, Marchetti2021}. Additionally, while only one \textcolor{black}{star unbound to the Galaxy} can be conclusively associated with an ejection from the Large Magellanic Cloud \citep{Przybilla2008HVS3, Erkal2019}, it is a potentially fruitful font of \textcolor{black}{unbound stars} in the future \citep{Boubert2016,Boubert2017,Evans2021}. 

It is important to ensure that any future \textit{Gaia} DR3 and/or DR4 HVS sample is free from contamination by unbound stars that were not ejected from the GC. To this end, we check whether the HVSs we predict to appear in \textit{Gaia} DR4 in our fiducial model can be confidently associated with an origin in the GC, or whether their origins will be ambiguous enough to warrant concern over contamination. 

\begin{figure*}
    \centering
    \includegraphics[width=2\columnwidth]{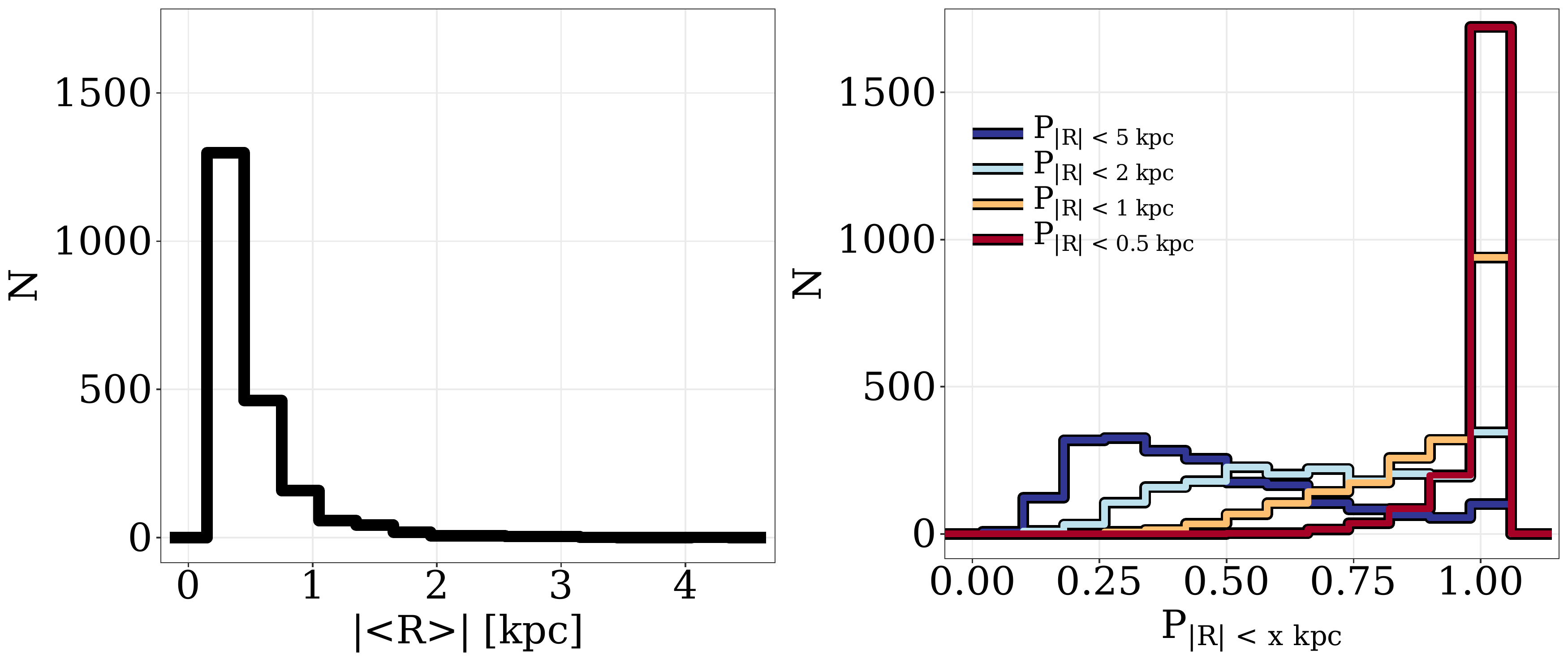}
    \caption{\textit{Left:} The distribution among 2137 fiducial-model DR4 HVSs of |<$R$>|, the distance between the GC and the average disc-crossing location of the HVS. \textit{Right}: The distribution of $P_{\rm R<x kpc}$, the probability that an HVS was ejected from within $x \, \mathrm{kpc}$ of the GC, where $x$ is $0.5 \, \mathrm{kpc}$ (dark blue), $1 \, \mathrm{kpc}$ (light blue), $2 \, \mathrm{kpc}$ (orange) and $5 \, \mathrm{kpc}$ (red).}
    \label{fig:Pdisk}
\end{figure*}

From 200 stacked simulations of our fiducial model, we obtain a sample of 2137 DR4 HVSs. For each star, we draw 1000 Monte Carlo (MC) realizations of the mock \textit{Gaia} parallax, proper motion and radial velocity observational uncertainties (see Sec. \ref{sec:methods:observations}), assuming uncorrelated Gaussian errors. For each of the 1000 realizations, we propagate the star backwards in time for $100 \, \mathrm{Myr}$ assuming the best-fit Galactic potential and solar position/velocity of \citet{McMillan2017}. In each case we record the disc-crossing location of the realization $R\equiv(x_0,y_0)$ in Galactic Cartesian coordinates when $z=0$. In the left panel of Fig. \ref{fig:Pdisk} we show the distribution among all the HVSs in our sample of |<$R$>|, i.e. the displacement of the average disc-crossing location from the GC. Reassuringly, the vast majority of DR4 HVSs track reliably back to the GC -- the disc-crossing centroids of 81\% of HVSs are within $1 \, \mathrm{kpc}$ of the GC, and 96\% of are within $2 \, \mathrm{kpc}$. We explore the spread of the disc-crossing locations in the right panel. The dark blue line shows the distribution of $P_{|R|<0.5 \, \mathrm{kpc}}$, i.e. the fraction of the MC realizations for each star which cross the Galactic disc within $0.5 \, \mathrm{kpc}$ of the GC. Only 25\% of stars have $P_{|R|<0.5 \, \mathrm{kpc}}$>50\%. Results improve considerably, however, using less restrictive cutoff radii -- half of all mock HVSs have $P_{|R|<2 \, \mathrm{kpc}}>0.8$ (orange line) and 87\% have $P_{|R|<5 \, \mathrm{kpc}}>0.8$ (red line).

To sum up, among the detected HVS candidates in \textit{Gaia} DR4, it will be straightforward to distinguish genuine GC-ejected HVSs from interloping unbound hyper-runaway stars or unbound stars of extragalactic origin. Upon propagating HVS candidates backwards in time, virtually all genuine HVSs will have average disc-crossing locations within $\sim$2$\, \mathrm{kpc}$ of the GC and a spread of possible disc-crossing locations contained to within $\sim$5$\, \mathrm{kpc}$ of the GC. We are therefore justified in claiming that contamination of any future HVS sample is not a worry.

\bsp	
\label{lastpage}
\end{document}